\documentclass[12pt]{JHEP3}
\usepackage{epsfig,amssymb,euscript}
\usepackage{amsmath,amscd}

\numberwithin{equation}{section}
\newcommand{\be}{\begin{eqnarray}}
\newcommand{\ee}{\end{eqnarray}}
\newcommand{\bea}{\begin{eqnarray}}
\newcommand{\eea}{\end{eqnarray}}
\newcommand{\ba}{\begin{array}}
\newcommand{\ea}{\end{array}}

\input{tcilatex}

\title{Half-Supersymmetric Solutions in Five-Dimensional Supergravity}

\author{Jan B. Gutowski \\ DAMTP, Centre for Mathematical Sciences\\
University of Cambridge\\
Wilberforce Road, Cambridge, CB3 0WA, UK\\ E-mail: \email{ J.B.Gutowski@damtp.cam.ac.uk}}

\author{Wafic Sabra \\ Centre for Advanced Mathematical Sciences and Physics Department\\
American University of Beirut\\ Lebanon \\ E-mail: \email{ws00@aub.edu.lb}}

\abstract{We present a systematic classification of half-supersymmetric solutions of
gauged $N=2$, $D=5$ supergravity coupled to an arbitrary number of abelian
vector multiplets for which at least one of the Killing spinors generate
a time-like Killing vector.}


\begin{document}

\section{Introduction}

There has been a considerable research effort in recent years devoted to the
analysis and study of solutions of supergravity theories in various
dimensions and in particular those obtained as low energy limits of
superstring and M-theory. This effort is motivated by the important role
that black holes and domain walls have played in some of the recent
developments that took place in superstring theory. These include the
conjectured equivalence between string theory on anti-de~Sitter (AdS) spaces
and certain superconformal gauge theories living on the boundary
\cite{maldacena} known as the AdS/CFT correspondence, the understanding of the
microscopic analysis of black hole entropy \cite{strominger} and the
understanding of various duality symmetries relating string theories to each
other and to M-theory. An interesting possibility that arises from the
conjectured AdS/CFT\ correspondence is the ability to obtain information of
the nonperturbative structure of field theories by studying dual classical
gravitational configurations. A notable example in this direction is the
Hawking-Page phase transition \cite{hawkpage} which was interpreted in
\cite{witten} as a thermal phase transition from a confining to a deconfining
phase in the dual $D=4$, $N=4$ super Yang-Mills theory. \ Various
interesting results using Anti-de Sitter black holes and their CFT duals
have been obtained in recent years (see for example \cite{chamb, empar,
cakl, awad, hawk, cvet}).

In this paper we will focus on the study of supersymmetric solutions in
five-dimensional $N=2$ gauged supergravity coupled to abelian vector
multiplets \cite{gunaydin}. These solutions are relevant for the holographic
descriptions of four dimensional field theories with less than maximal
supersymmetry. Explicit supersymmetric black holes for these theories were
constructed in \cite{electric}. However, these solutions have naked
singularities or naked closed time-like curves. Domain walls and magnetic
strings were also constructed in \cite{magnetic}. Obviously one would like
to study the general structure of supersymmetric solutions in five
dimensions rather than some specific solutions based on a certain ansatz.
The main purpose of this paper is to construct a systematic classification
of half-supersymmetric solutions.

The first systematic classification of supersymmetric solutions, following
the results of \cite{chris}, was performed in \cite{tod} \ for minimal $N=2$
supergravity in $D=4$. In \cite{tod} it was shown that supersymmetric
solutions fall into two classes which depend on whether the Killing vector
obtained from the Killing spinor is time-like or null. For the time-like
case, one obtains the Israel-Wilson-Perjes class of solutions and the null
solutions are pp-waves. Further generalizations were presented in
\cite{todagain}. More recently and motivated by the results of Tod, purely
bosonic supersymmetric solutions of minimal $N=2,$ $D=5$ were classified in
\cite{pakis}. The basic idea in this analysis is to assume the existence of
a Killing spinor, (i.e., to assume that the solution preserves at least one
supersymmetry) and construct differential forms as bilinears in the Killing
spinor. Then Fierz identities and the vanishing of the supersymmetry
transformation of the fermionic fields in a bosonic background provide a set
of algebraic and differential equations for the spinor bilinear differential
forms which can be used to deduce the form of the metric and gauge fields.
Such a general framework provides a powerful method for obtaining many new
solutions, in contrast to the earlier methods that start with an ansatz for
the metric and assume certain symmetries for the solution from the outset.
The strategy of \cite{pakis} was used later to perform similar
classifications of supersymmetric solutions in various supergravity
theories. In particular, in \cite{gaungut} the classification of $1/4$
supersymmetric solutions of the minimal gauged $N=2,$ $D=5$ supergravity was
performed.

Explicit supersymmetric asymptotically anti-de Sitter black hole solutions
with no closed time loops or naked singularities were constructed for the
minimal supergravity theory in \cite{gutowski1}. The results of
\cite{gaungut} for the time-like solutions were generalized in \cite{gutowski2}
to the non-minimal case where the scalar fields live on symmetric spaces and
explicit solutions for the $U(1)^{3}$ theory (with three $R$ -charges) were
also constructed. The constraint of symmetric spaces was relaxed in \cite{gs},
where solutions with a null Killing vector in both gauged and ungauged
theories were also obtained.

In this paper we focus on the classification of half supersymmetric
solutions in gauged $N=2,$ $D=5$ supergravity with vector multiplets. Half
supersymmetric solutions have two Killing spinors from which one can
construct two Killing vectors as bilinears in the Killing spinors. These
vectors could be either time-like or null. Therefore one has to consider
three cases depending on the nature of the Killing spinors and vectors
considered. In our present work we will focus on the cases where the
solutions contain at least one Killing spinor with an associated time-like
Killing vector. In order to investigate supersymmetric solutions with more
than one Killing spinor, it is very useful to express the Killing spinors in
terms of differential forms \cite{lawson}, \cite{wang}, \cite{harvey}. Such
a method, known now by the spinorial geometry method, has been very
efficient in classifying solutions of supergravity theories in ten and
eleven dimensions \cite{papadgran2005a}, \cite{papadgran2005b},
\cite{papadgran2006a} \cite{papadgran2006b}. The spinorial geometry method
has also been recently used to classify half-supersymmetric solutions
in $N=2$, $D=4$ supergravity \cite{roest2007}.

We organize our work as follows. In section~two, we present the basic
structure of the theory of $N=2$, $D=5$ gauged supergravity coupled to
abelian vector multiplets and the equations of motion. In section three we
express spinors in five dimensions as differential forms on
$\Lambda ^{\ast}(\mathbb{R}^{2})\otimes $ $\mathbb{C}$.
We start with the generic form of
the spinor and then use the gauge symmetries ($U(1)$ and $Spin(4,1))$
preserving the symplectic Majorana condition to write down two canonical
forms for a single symplectic Majorana spinor corresponding to time-like and
null Killing vectors. In section four, we derive the conditions for quarter
supersymmetric solutions with time-like Killing vector. In section five, the
$N=1$ Killing constraints, i. e., the conditions for a time-like quarter
supersymmetric solution, are then substituted into the generic Killing
spinor equations and the resulting equations are rewritten in the form of
constraints on the K\"{a}hler base. Section six contains a detailed
classification of half-supersymmetric solutions. Our paper ends with two
appendices. Appendix A deals with the determination of the linear system
obtained from the Killing spinor equations. Appendix B discusses the
integrability conditions of the Killing spinor equations. There it is
demonstrated that for a given background preserving at least half of the
supersymmetry, where at least one of the Killing spinors generates a
time-like Killing vector, all of the Einstein, gauge and scalar field
equations of motion hold automatically provided that the Bianchi identity is
satisfied.

\section{$N=2$ supergravity}

In this section, we review briefly some aspects of the $N=2$, $D=5$ gauged
supergravity coupled to abelian vector multiplets is~\cite{gunaydin}. The
bosonic action of the theory is
\begin{equation}
S={\frac{1}{16\pi G}}\int \left( -{}R+2\chi ^{2}{\mathcal{V}}\right)
{\mathcal{\ast }}1+Q_{IJ}\left( dX^{I}\wedge \star dX^{J}-F^{I}\wedge \ast
F^{J}\right) -{\frac{C_{IJK}}{6}}F^{I}\wedge F^{J}\wedge A^{K}
\label{action}
\end{equation}
where $I,J,K$ take values $1,\ldots ,n$ and $F^{I}=dA^{I}$ are the two-forms
representing gauge field strengths (one of the gauge fields corresponds to
the graviphoton). The metric has mostly negative signature. The constants
$C_{IJK}$ are symmetric in $IJK$ and are not assumed to satisfy the
non-linear \textquotedblleft adjoint identity\textquotedblright\ which
arises when the scalars lie in a symmetric space \cite{gunaydin}; though we
will assume that $Q_{IJ}$ is invertible, with inverse $Q^{IJ}$. The $X^{I}$
are scalar fields subject to the constraint
\begin{equation}
{\frac{1}{6}}C_{IJK}X^{I}X^{J}X^{K}=1\,.  \label{eqn:conda}
\end{equation}
The fields $X^{I}$ can thus be regarded as being functions of $n-1$
unconstrained scalars $\phi ^{r}$. It is convenient to define
\begin{equation}
X_{I}\equiv {\frac{1}{6}}C_{IJK}X^{J}X^{K}
\end{equation}
so that the condition~({\ref{eqn:conda}}) becomes
\begin{equation}
X_{I}X^{I}=1\,.
\end{equation}
In addition, the coupling $Q_{IJ}$ depends on the scalars via
\begin{equation}
Q_{IJ}={\frac{9}{2}}X_{I}X_{J}-{\frac{1}{2}}C_{IJK}X^{K}
\end{equation}
so in particular

\bigskip

\begin{equation}
Q_{IJ}X^{J}={\frac{3}{2}}X_{I}\,,\qquad Q_{IJ}\partial _{r}X^{J}=-{\frac{3}
{2}}\partial _{r}X_{I}\,.
\end{equation}
where $\partial _{r}$ denotes differentiation with respect to $\phi ^{r}.$
The scalar potential can be written as
\begin{equation}
{\mathcal{V}}=9V_{I}V_{J}(X^{I}X^{J}-{\frac{1}{2}}Q^{IJ})
\end{equation}
where $V_{I}$ are constants.

Bosonic backgrounds are said to be supersymmetric if there exists a spinor
$\epsilon ^{a}$ for which the supersymmetry variations of the gravitino and
dilatino vanish in the given background. For the gravitino this requires

\begin{equation}
\left[ \nabla _{\mu }+\frac{1}{8}X_{I}\left( \gamma _{\mu }F^{I}{}_{\rho
\sigma }\gamma ^{\rho \sigma }-{6}F^{I}{}_{\mu \rho }\gamma ^{\rho }\right)
\right] \epsilon ^{a}-{\frac{\chi }{2}}V_{I}(X^{I} \gamma
_{\mu}-3A^{I}{}_{\mu })\epsilon ^{ab}\epsilon ^{b}=0,  \label{eqn:grav}
\end{equation}
and for the dilatino it requires

\begin{equation}
\left[ {\frac{1}{4}}\left( Q_{IJ}\gamma ^{\mu \nu }F^{J}{}_{\mu \nu }+{3}
\gamma ^{\mu }\nabla _{\mu }X_{I}\right) \epsilon ^{a}-{\frac{3\chi }{2}}
V_{I}\epsilon ^{ab}\epsilon ^{b}\right] {\partial _{r}X^{I}}=0\,.
\label{eqn:dil}
\end{equation}
The Einstein equation derived from (\ref{action}) is given by

\begin{equation}
{}R_{\alpha \beta }+Q_{IJ}\left( F^{I}{}_{\alpha \lambda }
F^{J}{}_{\beta}{}^{\lambda }-\nabla _{\alpha }X^{I}\nabla _{\beta }X^{J} -
{\frac{1}{6}}g_{\alpha \beta }F^{I}{}_{\mu \nu }F^{J\mu \nu }\right)
-{\frac{2}{3}}g_{\alpha \beta }\chi ^{2}\mathcal{V}=0\,.  \label{eqn:ein}
\end{equation}
The Maxwell equations (varying $A^{I}$) are

\begin{equation}
d\left( Q_{IJ}\star F^{J}\right) =-{\frac{1}{4}}C_{IJK}F^{J}\wedge F^{K}\,.
\label{eqn:gauge}
\end{equation}
The scalar equations (varying $\phi ^{r}$) are

\begin{eqnarray}
&&\bigg[{-}d(\star dX_{I})+\left( X_{M}X^{P}C_{NPI}-{\frac{1}{6}}
C_{MNI}\right) (F^{M}\wedge \star F^{N}-dX^{M}\wedge \star dX^{N})  \notag \\
&&-{\frac{3}{2}}\chi ^{2}V_{M}V_{N}Q^{ML}Q^{NP}C_{LPI}\mathrm{dvol}\bigg]
{\partial _{r}X^{I}}=0\,.\qquad
\end{eqnarray}
If a quantity $L_{I}$ satisfies $L_{I}\partial _{r}X^{I}=0,$ then there must
be a function $\Upsilon$ such that $L_{I}=\Upsilon X_{I}$. This implies that
the dilatino equation~({\ref{eqn:dil}}) can be simplified to

\begin{equation}
F^{I}{}_{\mu \nu }\gamma ^{\mu \nu }\epsilon ^{a}=\left(
X^{I}X_{J}F^{J}{}_{\mu \nu }\gamma ^{\mu \nu }+2\gamma ^{\mu }\nabla _{\mu
}X^{I}\right) \epsilon ^{a}-4\chi V_{J}(X^{I}X^{J}-{\frac{3}{2}}
Q^{IJ})\epsilon ^{ab}\epsilon ^{b}  \label{eqn:newdil}
\end{equation}
and the scalar equation can be written as

\begin{eqnarray}
&-&d\left( \star dX_{I}\right) +\left( {\frac{1}{6}}C_{MNI}-{\frac{1}{2}}
X_{I}C_{MNJ}X^{J}\right) dX^{M}\wedge \star dX^{N}  \notag \\
&&+\left( X_{M}X^{P}C_{NPI}-{\frac{1}{6}}C_{MNI}-6X_{I}X_{M}X_{N}
+{\frac{1}{6}}X_{I}C_{MNJ}X^{J}\right) F^{M}\wedge \star F^{N}  \notag \\
&+&3\chi ^{2}V_{M}V_{N}\left( {\frac{1}{2}}
Q^{ML}Q^{NP}C_{LPI}+X_{I}Q^{MN}-2X_{I}X^{M}X^{N}\right) \mathrm{dvol}=0\ .
\end{eqnarray}

\section{Spinors in Five Dimensions}

Following \cite{lawson, wang, harvey} , we write spinors in five dimensions
as forms on $\Lambda^{\ast }(\mathbb{R}^{2})\otimes \mathbb{C}$. We
represent a generic spinor $\eta $ in the form

\begin{equation}
\eta =\lambda 1+\mu^{i}e^{i}+\sigma e^{12},
\end{equation}
where $e^{1}$, $e^{2}$ are 1-forms on $\mathbb{R}^{2}$, $e^{12}=e^{1}\wedge
e^{2}$ and $\lambda ,\mu^{i}$ and $\sigma $ are complex functions.

The action of $\gamma $-matrices on these forms is given by

\begin{equation}
\gamma_{i}=i(e^{i}\wedge +i_{e^{i}}),\text{ \ \ \ \ } \gamma_{i+2}=-e^{i}
\wedge +i_{e^{i}}.
\end{equation}
We define $\gamma_{0}$ by $\gamma_{0}=\gamma_{1234}.$ This satisfies

\begin{equation}
\gamma_{0}1=1,\quad \gamma_{0}e^{12}=e^{12},\quad \gamma_{0}e^{i}=-e^{i}\ .
\end{equation}
The charge conjugation operator $C$ is defined by

\begin{equation}
C1=-e^{12},\quad Ce^{12}=1,\quad Ce^{i}=-\epsilon _{ij}e^{j}\
\end{equation}
where $\epsilon _{ij}=\epsilon ^{ij}$ is antisymmetric with $\epsilon
_{12}=1 $.

The Killing spinors $\epsilon^a$ of the theory satisfy a symplectic Majorana
constraint which is

\begin{equation}
(\epsilon^{a})^{\ast }=\epsilon^{ab}\gamma_{0}C\epsilon^{b}  \label{macon}
\end{equation}
so if one writes

\begin{equation}
\epsilon^{1}=\lambda 1+\mu^{i}e^{i}+\sigma e^{12},  \label{maj1}
\end{equation}
then $\epsilon ^{2}$ is fixed via

\begin{equation}
\epsilon^{2}=-\sigma^{\ast }1-\epsilon_{ij} (\mu
^{i})^{\ast}e^{j}+\lambda^{\ast }e^{12}.  \label{maj2}
\end{equation}
We note the useful identity

\begin{equation}
(\gamma _{\mu})^{\ast }=-\gamma _{0}C\gamma _{\mu}\gamma _{0}C \ .
\end{equation}
It will be particularly useful in our work to complexify the gamma
operators. Therefore we write

\begin{eqnarray}
\gamma _{p} &=&{\frac{1}{\sqrt{2}}}(\gamma _{p}-i\gamma _{p+2})=\sqrt{2}
ie^{p}\wedge  \notag \\
\gamma _{\bar{p}} &=&{\frac{1}{\sqrt{2}}}(\gamma _{p}+i\gamma _{p+2})=
\sqrt{2}ii_{e^{p}}.
\end{eqnarray}

\subsection{Gauge transformations and $N=1$ spinors}

There are two types of gauge transformation that preserve the symplectic
Majorana condition (\ref{macon}). First, we have the $U(1)$ gauge
transformations described by

\begin{equation}
\left(
\begin{array}{c}
\epsilon ^{1} \\
\epsilon ^{2}
\end{array}
\right) =\left(
\begin{array}{cc}
\cos \theta & \sin \theta \\
-\sin \theta & \cos \theta
\end{array}
\right) \left(
\begin{array}{c}
\epsilon ^{1} \\
\epsilon ^{2}
\end{array}
\right)  \label{u1gauge}
\end{equation}
and there are also $Spin(4,1)$ gauge transformations of the form

\begin{equation}
\epsilon ^{a}\rightarrow e^{{\frac{1}{2}}f^{\mu \nu}\gamma _{\mu
\nu}}\epsilon ^{a},
\end{equation}
for real functions $f^{\mu \nu}$.

Note in particular that ${\frac{1 }{2}}(\gamma_{12}+\gamma_{34})$, ${\frac{1
}{2}}(\gamma_{13}-\gamma_{24})$ and ${\frac{1 }{2}}(\gamma_{14}+\gamma_{23})$
generate a $SU(2)$ which leaves $1$ and $e^{12}$ invariant and acts on $e^1$
, $e^2$; whereas ${\frac{1 }{2}}(\gamma_{12}-\gamma_{34})$, ${\frac{1 }{2}}
(\gamma_{13}+\gamma_{24})$ and ${\frac{1 }{2}}(\gamma_{14}-\gamma_{23})$
generate another $SU(2)$ which leaves the $e^i$ invariant but acts on $1$
and $e^{12}$. In addition, $\gamma_{03}$ generates a $SO(1,1)$ which acts
(simultaneously) on $1, e^1$ and $e^2, e^{12}$ , whereas $\gamma_{04}$
generates another $SO(1,1)$ which acts (simultaneously) on $1, e^2$ and
$e^1, e^{12}$.

\bigskip Therefore, for a single symplectic Majorana spinor, one can always
use $Spin(4,1)$ gauge transformations to write

\begin{equation}
\epsilon^{1}=f1,\quad \epsilon^{2}=fe^{12},  \label{eqn:spin1}
\end{equation}

or

\begin{equation}
\epsilon^{1}=fe^{1},\quad \epsilon^{2}=-fe^{2},  \label{eqn:spin2}
\end{equation}

or

\begin{equation}
\epsilon^{1}=f(1+e^{1}),\quad \epsilon^{2}=f(-e^{2}+e^{12}),
\label{eqn:spin3}
\end{equation}
for some real function $f$. \ However, under the transformation

\begin{eqnarray}
\epsilon ^{a} &\rightarrow &\gamma_{1}\epsilon ^{a},  \notag \\
\gamma_{\mu} &\rightarrow &-\gamma_{1}\gamma_{\mu}\gamma_{1},  \notag \\
C &\rightarrow &-\gamma_{1}C\gamma_{1},
\end{eqnarray}
the spinor in ({\ref{eqn:spin2}}) transforms as

\begin{equation}
\epsilon^{1}\rightarrow if1,\quad \epsilon^{2}\rightarrow -ife^{12}
\label{eqn:next1}
\end{equation}
and

\begin{equation}
\gamma_{0}\rightarrow -\gamma_{0},\quad \gamma_{1}\rightarrow \gamma_{1},
\quad \gamma_{2}\rightarrow -\gamma_{2},\quad \gamma_{3}\rightarrow
-\gamma_{3},\quad \gamma_{4}\rightarrow -\gamma_{4}
\end{equation}
and $C$ is unchanged. This transformation corresponds to reflections in the
$0,2,3,4$ directions. Moreover, the spinor in ({\ref{eqn:next1}}) is
equivalent to that in ({\ref{eqn:spin1}}) under a $SU(2)$ gauge
transformation. The spinors corresponding to ({\ref{eqn:spin1}}) and
({\ref{eqn:spin2}}) are therefore equivalent under these transformations. Hence,
for a single spinor, one need only consider the cases ({\ref{eqn:spin1}})
and ({\ref{eqn:spin3}}).

\subsection{Differential Forms from Spinors}

In order to define differential forms, we first define a Hermitian inner
product on $\Lambda^{\ast }\mathbb{R}^{2}\otimes \mathbb{C}$ by

\begin{equation}
\langle z^0 1 + z^1 e^1 + z^2 e^2 + z^3 e^{12}, w^0 1 + w^1 e^1 + w^2 e^2 +
w^3 e^{12}\rangle = \sum_{\alpha=0}^3 {\bar{z}}^{\alpha }w^{\alpha } \ .
\end{equation}
Then $Spin(4,1)$ gauge-invariant $k$-forms are obtained from spinors
$\epsilon$, $\eta$ via

\begin{equation}
\alpha (\epsilon ,\eta )_{\mu_{1},\dots ,\mu_{k}}=-\langle C\epsilon^{\ast
}, \gamma _{\mu_{1},\dots ,\mu_{k}}\eta \rangle .
\end{equation}

In particular, for the generic Majorana spinor given in ({\ref{maj1}}) and
({\ref{maj2}}) one finds

\begin{equation}
\alpha (\epsilon^{a},\epsilon^{b})_{\mu _{1},\dots ,\mu _{k}}=\langle
\epsilon^{ac}\gamma_{0}\epsilon^{c}, \gamma _{\mu_{1},\dots
,\mu_{k}}\epsilon^{b}\rangle .
\end{equation}
The scalars are then given by

\begin{equation}
\alpha (\epsilon ^{a},\epsilon ^{b})=\epsilon ^{ab}(|\sigma |^{2}+|\lambda
|^{2}-|\mu _{1}|^{2}-|\mu _{2}|^{2}).
\end{equation}
Hence, by comparing with \cite{pakis}, it is clear that the spinor given in
({\ref{eqn:spin1}}) corresponds to the time-like class of solutions, whereas
that in ({\ref{eqn:spin3}}) is in the null class of solutions. With a slight
abuse of notation, we shall refer to the corresponding Killing spinors as
being either time-like or null.

\subsection{Canonical $N=2$ spinors}

We will now assume that there are two linearly independent symplectic
Majorana Killing spinors $\epsilon ^{a}$, $\eta ^{a}$, where $\epsilon ^{a}$
is time-like. In appendix B it is demonstrated that the existence of such
spinors is sufficient to ensure that the scalar, gauge and Einstein
equations of motion hold automatically from the integrability conditions,
provided one assumes that the Bianchi identities are satisfied. So the only
equations which must be solved are the Killing spinor equations together
with the Bianchi identity.

{}From the previous reasoning, we can take $\epsilon^a$ to have the
canonical form.

\begin{equation}
\epsilon ^{1}=f,\text{ \ \ }\epsilon ^{2}=fe^{12}  \label{next2}
\end{equation}
for $f\in \mathbb{R}$. Next consider $\eta ^{a}$ given by

\begin{eqnarray}
\eta ^{1} &=&\lambda 1+\mu ^{i}e^{i}+\sigma e^{12},  \label{sali} \\
\eta ^{2} &=&-\sigma ^{\ast }1-\epsilon _{ij}(\mu ^{i})^{\ast }e^{j}+\lambda
^{\ast }e^{12}  \label{secons}
\end{eqnarray}
for complex $\lambda ,\mu _{i},\sigma $. It is possible to simplify $\eta
^{a}$ a little using gauge transformations which leave $\epsilon ^{a}$
invariant. In particular, by using an appropriate $SU(2)$ transformation,
one could for example set $\mu ^{2}=0$ with $\mu ^{1}\in \mathbb{R}$.
However, we will not make this gauge choice.

\subsection{The 1/4 Supersymmetric time-like Solution}

In this section we obtain the time-like solutions preserving a quarter of
the supersymmetry using the spinorial geometry method. These solutions were
derived in \cite{gutowski2, gs}. In order to obtain 1/4 supersymmetric
solutions with time-like Killing spinor, it suffices to consider the
equations ({\ref{alg}})-({\ref{algg}}) and set $\sigma =\mu ^{p}=0$ and
$\lambda =f$. Then from the dilatino equation, we find
\begin{eqnarray}
F^{I}{}_{m}{}^{m} &=&X^{I}H_{m}{}^{m}-\partial _{0}X^{I},  \label{dione} \\
F^{I}{}_{0n} &=&X^{I}H_{0n}-\partial _{n}X^{I}, \\
\left( F^{I}{}_{mn}-X^{I}H_{mn}\right) \epsilon ^{mn} &=&2\chi
V_{J}(X^{I}X^{J}-{\frac{3}{2}}Q^{IJ}),  \label{dithree}
\end{eqnarray}
whereas from the gravitino equation we find

\begin{eqnarray}
\frac{1}{f}\partial _{0}f-\frac{1}{4}\left( {2}\omega
_{0,m}{}^{m}+H_{m}{}^{m}\right) &=&0,  \label{equ1} \\
\omega _{0,0n}+H_{0n} &=&0,  \label{equ2} \\
\left( H_{mn}+2\omega _{0,mn}\right) \epsilon ^{mn}+2\chi
V_{I}(X^{I}-3A^{I}{}_{0}) &=&0,  \label{equ3} \\
\frac{1}{f}\partial _{p}f-\frac{1}{4}\left( {2}\omega
_{p,m}{}^{m}-3H_{0p}\right) &=&0,  \label{equ4} \\
H_{p\bar{q}}-\frac{1}{3}\left( H_{m}{}^{m}\delta _{p\bar{q}}-
2\omega _{p,0\bar{q}}\right) &=&0,  \label{equ5} \\
-\omega _{p,\bar{m}\bar{n}}\epsilon ^{\bar{m}\bar{n}}+H_{0\bar{n}}
\epsilon ^{\bar{n}}{}_{p}+3\chi V_{I}A^{I}{}_{p} &=&0,  \label{equ6} \\
\frac{1}{f}\partial _{\bar{p}}f-{\frac{1}{4}}\left(
{2}\omega _{\bar{p},m}{}^{m}-H_{0\bar{p}}\right) &=&0,  \label{equ7} \\
\omega _{p,0q}+\left( {\frac{1}{4}}H_{mn}\epsilon ^{mn}-\chi
V_{I}X^{I}\right) \epsilon _{pq} &=&0,  \label{equ8} \\
3\chi V_{I}A^{I}{}_{\bar{p}}-\omega _{\bar{p},\bar{m}\bar{n}}
\epsilon ^{\bar{m}\bar{n}} &=&0.  \label{equ9}
\end{eqnarray}
To analyze this linear system, we will first consider the gravitino
equations. Note that ({\ref{equ1}}) implies that

\begin{equation}
\partial_{0}f=0,  \label{c1}
\end{equation}
and
\begin{equation}
H_{m}{}^{m}=-2\omega_{0,m}{}^{m}.  \label{c2}
\end{equation}
Next, consider ({\ref{equ2}}), ({\ref{equ4}}) and ({\ref{equ7}}). These imply

\begin{eqnarray}
H_{0p} &=&-\frac{2}{f}\partial _{p}f,  \label{c3} \\
\omega_{0,0p} &=&\frac{2}{f}\partial_{p}f,  \label{c4} \\
\omega_{p,m}{}^{m} &=&-\frac{1}{f}\partial_{p}f.  \label{c5}
\end{eqnarray}
From ({\ref{equ3}}) and ({\ref{equ8}}) we find

\begin{equation}
\omega_{(p,q)0}=0,  \label{c6}
\end{equation}
and

\bigskip

\begin{equation}
\omega _{\lbrack {\bar{m}},0]{\bar{n}}}\epsilon ^{\bar{m}\bar{n}}+{\frac{%
3\chi }{2}}V_{I}(A^{I}{}_{0}-X^{I})=0.  \label{c7}
\end{equation}%
{}From ({\ref{equ5}}) we find

\begin{equation}
\omega_{(p,{\bar{q}})0}=0  \label{c8}
\end{equation}
and from ({\ref{equ6}}) and ({\ref{equ9}}) we obtain

\begin{equation}
\omega _{p,mn}\epsilon^{mn}-\omega_{p,\bar{m}\bar{n}} \epsilon^{\bar{m}\bar{n%
}}-\frac{2}{f}\epsilon^{\bar{n}}{}_{p}\partial_{\bar{n}}f=0  \label{c9}
\end{equation}
and

\begin{equation}
3\chi V_{I}A^{I}{}_{p}=\omega _{p,mn}\epsilon^{mn}.  \label{c10}
\end{equation}
Hence, to summarize, we obtain the following purely geometric constraints

\begin{eqnarray}
\partial_{0}f &=&0,  \notag \\
\omega_{0,0p} &=&2\frac{\partial _{p}f}{f},  \notag \\
\omega_{(p,q)0} &=&0,  \notag \\
\omega_{(p,{\bar{q}})0} &=&0,  \label{kill1}
\end{eqnarray}
together with

\begin{eqnarray}
\omega _{p,m}{}^{m}+\frac{\partial _{p}f}{f} &=&0,  \notag \\
\omega _{p,mn}\epsilon ^{mn}-\omega _{p,\bar{m}\bar{n}}\epsilon ^{\bar{m}%
\bar{n}}-\frac{2}{f}\epsilon {}^{\bar{n}}{}_{p}\partial _{\bar{n}}f &=&0.
\label{kahlr}
\end{eqnarray}%
It is straightforward to show that the constraints ({\ref{kill1}}) are the
necessary and sufficient conditions for the 1-form
\begin{equation}
\kappa =f^{2}\mathbf{e}^{0}  \label{c12}
\end{equation}%
to define a Killing vector $V$. This form is, as expected, the 1-form spinor
bilinear which is obtained from $\epsilon ^{a}$. Note that this Killing
vector satisfies
\begin{equation}
\mathcal{L}_{V}\mathbf{e}^{0}=0\ .
\end{equation}%
In fact, one can choose a gauge in which the Lie derivative of the vielbein
with respect to $V$ vanishes. To see this, note that
\begin{equation}
\mathcal{L}_{V}\mathbf{e}^{p}=f^{2}(\omega _{0,\bar{p}q}-\omega _{q,\bar{p}%
0})\mathbf{e}^{q}+f^{2}\epsilon _{\bar{p}\bar{q}}\omega _{\lbrack 0,\bar{m}]%
\bar{n}}\epsilon ^{\bar{m}\bar{n}}\mathbf{e}^{\bar{q}}.
\end{equation}%
From ({\ref{c7}}) we note that $\omega _{\lbrack 0,\bar{m}]\bar{n}}\epsilon
^{\bar{m}\bar{n}}\in \mathbb{R}$. Without loss of generality we can make a $%
U(1)$ gauge transformation in order to set
\begin{equation}
V_{I}A_{0}^{I}=V_{I}X^{I}\ .
\end{equation}%
This gauge transformation alters the form of the Killing spinors via the
transformation given in ({\ref{u1gauge}}). However, the Killing spinors can
be restored to their original form by making a $U(1)\in SU(2)\subset
Spin(4,1)$ gauge transformation generated by $\gamma _{12}-\gamma _{34}$.
Working in this gauge, ({\ref{c7}}) implies that $\omega _{\lbrack 0,\bar{m}]%
\bar{n}}\epsilon ^{\bar{m}\bar{n}}=0$ and hence
\begin{equation}
\mathcal{L}_{V}\mathbf{e}^{p}=A^{p}{}_{q}\mathbf{e}^{q}
\end{equation}%
where the constraints in ({\ref{equ5}}), ({\ref{c2}}) and ({\ref{kill1}})
imply that $A\in su(2)$. By making a further $SU(2)\subset Spin(4,1)$ gauge
transformation generated by $\gamma _{12}+\gamma _{34},\gamma _{13}-\gamma
_{24},\gamma _{14}+\gamma _{23}$ which leaves $1,e^{12}$ invariant and maps $%
e^{p}\rightarrow X^{p}{}_{q}e^{q}$ for $X\in SU(2)$, we can without loss of
generality take $A=0$. In this basis the vielbein is time-independent.

Equation ({\ref{kahlr}}) also has a simple geometric interpretation. First
note that the only $U(1)$ gauge-invariant 2-form which can be obtained from $%
\epsilon $ is the real part of the 2-form
\begin{equation}
\alpha _{\mu \nu }(\epsilon ^{1},\epsilon ^{1})=-\langle Cf.1,\gamma _{\mu
\nu }f1\rangle =f^{2}\langle e^{12},\gamma _{\mu \nu }1\rangle .
\end{equation}%
The real part of this form (denoted by $J$) is then given by

\begin{equation}
J_{pq}=-f^{2}\epsilon _{pq},\quad J_{\bar{p}\bar{q}}=-f^{2} \epsilon _{\bar{p%
}\bar{q}}.
\end{equation}
It is convenient to make a conformal rescaling of the complexified basis and
define

\begin{equation}
\mathbf{e}^{p}=f^{-1}\mathbf{\hat{e}}^{p},\quad \mathbf{e}^{\bar{p}}=f^{-1}%
\mathbf{\hat{e}}^{\bar{p}}.
\end{equation}%
We shall refer to the 4-manifold with metric

\begin{equation}
{\hat{ds}}_{4}{}^{2}=2\left( \mathbf{\hat{e}}^{1}\mathbf{\hat{e}}^{\bar{1}}+%
\mathbf{\hat{e}}^{2}\mathbf{\hat{e}}^{\bar{2}}\right)  \label{confmet}
\end{equation}%
as the base space $B$. Then it is clear that $J$ defines an almost complex
structure on this 4-manifold. In fact, ({\ref{kahlr}}) implies that $J$ is
covariantly constant with respect to the Levi-civita connection of the base
manifold, and hence $B$ is a K\"{a}hler manifold (as expected) with K\"{a}%
hler form $J$.

Also, from ({\ref{equ9}}) we have

\begin{equation}
3\chi V_{I}A^{I}{}_{p}=\omega _{p,mn}\epsilon ^{mn}.  \label{c16}
\end{equation}
We remark that ({\ref{c16}}) implies that

\begin{equation}
{\mathcal{P=}}3\chi V_{I}(A^{I}{}_{p}\mathbf{e}^{p}+A^{I}{}_{\bar{p}}\mathbf{%
e}^{\bar{p}}).  \label{eqn:killpot1}
\end{equation}%
where ${\mathcal{P}}$ is (locally) the potential for the Ricci form of the K%
\"{a}hler base $B$ \footnote{%
If we are in the ungauged theory with $\chi =0$, then the vanishing of $%
\omega _{p,mn}$ is then sufficient to imply that the base $B$ is hyper-K\"{a}%
hler. But we shall take $\chi \neq 0$ throughout.}.

The remaining constraints on the $H$-flux are then

\begin{eqnarray}
H_{0p} &=&-2\frac{\partial _{p}f}{f},  \label{c17} \\
H_{mn}\epsilon ^{mn} &=&-2\omega _{0,mn}\epsilon ^{mn}+4\chi V_{I}X^{I}, \\
H_{p\bar{q}} &=&-\frac{2}{3}\left( \omega _{0,m}{}^{m}\delta _{p\bar{q}%
}+\omega _{0,p\bar{q}}\right) .
\end{eqnarray}

Finally, we substitute these constraints into the dilatino equations. From ({%
\ref{dione}})-(\ref{dithree}), we find

\begin{eqnarray}
\partial _{0}X^{I} &=&0, \\
F^{I}{}_{m}{}^{m} &=&-2\omega _{0,m}{}^{m}X^{I}, \\
F^{I}{}_{0n} &=&-\frac{1}{f^{2}}\partial _{n}(f^{2}X^{I}), \\
\left( F^{I}{}_{mn}+2X^{I}\omega _{0,mn}\right) \epsilon ^{mn} &=&3\chi
V_{J}(2X^{I}X^{J}-Q^{IJ}).  \label{field}
\end{eqnarray}

\section{Killing spinor in $N=1$ background}

In this section, we substitute the constraints obtained in the previous
section back into the generic Killing spinor equation ({\ref{alg}})-({\ref%
{algg}}) and simplify as much as possible. We find from the dilatino
equation:

\begin{eqnarray}
\mu ^{m}\partial _{m}X^{I} &=&-\sqrt{2}\chi V_{J}(X^{I}X^{J}-{\frac{3}{2}}%
Q^{IJ})\func{Im}\sigma ,  \label{simp1} \\
\mu ^{m}\left[ F^{I}{}_{m\bar{q}}+{\frac{2}{3}}(\omega _{0,p}{}^{p}\delta _{m%
\bar{q}}+\omega _{0,m\bar{q}})X^{I}\right] &=&\chi V_{J}(X^{I}X^{J}-{\frac{3%
}{2}}Q^{IJ})\epsilon _{\bar{m}\bar{q}}(\mu ^{m})^{\ast },  \label{simp2} \\
\mu ^{m}\epsilon ^{\bar{n}}{}_{m}\partial _{\bar{n}}X^{I} &=&\sqrt{2}\chi
V_{J}(X^{I}X^{J}-{\frac{3}{2}}Q^{IJ})\func{Im}\lambda .  \label{simp3}
\end{eqnarray}%
And from the gravitino equations we find

\begin{eqnarray}
\partial _{0}\lambda &=&2i\left( \sqrt{2}\frac{\mu ^{m}}{f}\partial
_{m}f-\chi V_{I}X^{I}\func{Im}\sigma \right) ,  \label{simp4} \\
\partial _{0}\sigma &=&2i\left( \sqrt{2}\frac{\mu ^{m}}{f}\epsilon ^{\bar{n}%
}{}_{m}\partial _{\bar{n}}f+\chi V_{I}X^{I}\func{Im}\lambda \right) ,
\label{simp6} \\
\partial _{0}\mu _{\bar{q}} &=&-\frac{2}{3}\mu ^{m}\left( {2}\omega _{0,m%
\bar{q}}-\omega _{0,p}{}^{p}\delta _{m\bar{q}}\right) +2\chi
V_{I}X^{I}\epsilon _{\bar{m}\bar{q}}(\mu ^{m})^{\ast },  \label{simp5}
\end{eqnarray}

and

\begin{eqnarray}
\partial _{\bar{p}}(\frac{\sigma }{f}) &=&\frac{i}{f}\left( -\sqrt{2}\mu _{%
\bar{p}}(\omega _{0,\bar{r}\bar{n}}\epsilon ^{\bar{r}\bar{n}}-{\frac{3\chi
V_{I}X^{I}}{2}})+\omega _{\bar{p},\bar{m}\bar{n}}\epsilon ^{\bar{m}\bar{n}}%
\func{Im}\lambda \right) ,  \label{simp12} \\
\partial _{p}(\frac{\lambda }{f}) &=&\frac{i}{f}\left( \sqrt{2}\mu
^{m}(2\omega _{0,pm}-{\frac{3\chi V_{I}X^{I}}{2}}\epsilon _{pm})-\omega
_{p,mn}\epsilon ^{mn}\func{Im}\sigma \right) ,  \label{simp7} \\
\partial _{p}(\frac{\sigma }{f}) &=&{\frac{2i\sqrt{2}\mu ^{m}}{3f}}(\omega
_{0,p\bar{n}}\epsilon ^{\bar{n}}{}_{m}+\omega _{0,n}{}^{n}\epsilon _{pm})-{%
\frac{i\chi V_{I}X^{I}}{\sqrt{2}f}}(\mu _{\bar{p}})^{\ast }  \notag \\
&&+\frac{i}{f}\omega _{p,mn}\epsilon ^{mn}\func{Im}\lambda ,  \label{simp9}
\\
\partial _{\bar{p}}(\frac{\lambda }{f}) &=&{\frac{2i\sqrt{2}\mu ^{m}}{3f}}%
(\omega _{0,\bar{p}m}-\omega _{0,n}{}^{n}\delta _{\bar{p}m})+{\frac{i\chi
V_{I}X^{I}}{\sqrt{2}f}}\epsilon _{\bar{p}\bar{m}}(\mu ^{m})^{\ast }  \notag
\\
&&-\frac{i}{f}\omega _{\bar{p},\bar{m}\bar{n}}\epsilon ^{\bar{m}\bar{n}}%
\func{Im}\sigma ,  \label{simp10}
\end{eqnarray}

\begin{eqnarray}
\partial _{p}\mu _{\bar{q}} &=&(\mu ^{m})^{\ast }\left( \omega _{p,\bar{m}%
\bar{q}}+\epsilon ^{\bar{n}}{}_{p}\epsilon _{\bar{m}\bar{q}}\frac{\partial _{%
\bar{n}}f}{f}\right) -\mu ^{m}\left( 2\delta _{m\bar{q}}\frac{\partial _{p}f%
}{f}-\delta _{p\bar{q}}\frac{\partial _{m}f}{f}+\omega _{p,m\bar{q}}\right)
\notag \\
&&-\sqrt{2}\chi V_{I}X^{I}\delta _{p\bar{q}}\func{Im}\sigma ,  \label{simp8}
\\
\partial _{\bar{p}}\mu _{\bar{q}} &=&-\mu ^{m}\left( \delta _{\bar{q}m}\frac{%
\partial _{\bar{p}}f}{f}-\epsilon ^{\bar{n}}{}_{m}\epsilon _{\bar{p}\bar{q}}%
\frac{\partial _{\bar{n}}f}{f}+\omega _{\bar{p},m\bar{q}}\right) +\omega _{%
\bar{p},\bar{m}\bar{q}}(\mu ^{m})^{\ast }  \notag \\
&&+\sqrt{2}\chi V_{I}X^{I}\epsilon _{\bar{p}\bar{q}}\func{Im}\lambda .
\label{simp11}
\end{eqnarray}

Note that these equations admit a solution of the form $\lambda =\rho _{1}f$%
, $\sigma =\rho _{2}f$, $\mu ^{p}=0$ with $\rho _{1}$, $\rho _{2}$ real
constants, and no additional constraints on the fluxes or geometry. Hence we
observe that the generic time-like solution preserves $1/4$ supersymmetry.
More generally, if $\eta ^{1}$, $\eta ^{2}$ are symplectic Majorana Killing
spinors, then so are

\begin{equation}
(\eta ^{1})^{\prime }=\eta ^{2},\qquad (\eta ^{2})^{\prime }=-\eta ^{1}
\end{equation}
which is just a special case of ({\ref{u1gauge}}) with $\theta ={\frac{\pi }{%
2}}$. In particular, the equations computed above are invariant under the
transformations

\begin{eqnarray}
\lambda &\rightarrow &-\sigma ^{\ast },  \notag \\
\sigma &\rightarrow &\lambda ^{\ast },  \notag \\
\mu ^{p} &\rightarrow &\epsilon ^{p}{}_{\bar{q}}(\mu ^{q})^{\ast }
\end{eqnarray}
therefore it is clear that the Killing spinors arise in pairs.

\subsection{Solutions with $\protect\mu^p=0$}

Suppose we consider the case when $\mu ^{p}=0$. Then, assuming that $%
V_{I}X^{I}\neq 0$, we find from ({\ref{simp8}}) and ({\ref{simp11}}) that
\begin{equation}
\func{Im}\sigma =\func{Im}\lambda =0.  \label{imvanish}
\end{equation}%
If, however, $V_{I}X^{I}=0$ then ({\ref{simp1}}) and ({\ref{simp3}}) again
imply ({\ref{imvanish}}), as we assume that not all of the $V_{I}$ vanish.

Hence, from ({\ref{simp4}}), ({\ref{simp6}}), ({\ref{simp7}}) and ({\ref%
{simp9}}) it follows directly that $\lambda =\rho _{1}f$, $\sigma =\rho
_{2}f $, $\mu ^{p}=0$ with $\rho _{1}$, $\rho _{2}$ real constants. The
solution is therefore only 1/4-supersymmetric. Thus, to find new solutions
with enhanced supersymmetry, one must take $\mu ^{p}\neq 0$; henceforth we
shall assume that $\mu ^{p}\neq 0$.

\subsection{Constraints on the base space}

It will be particularly useful to rewrite the equations ({\ref{simp1}})-({%
\ref{simp11}}) in terms of constraints on the K\"{a}hler base. Throughout
this section, unless stated otherwise, tensor indices are evaluated with
respect to the 4-dimensional complex basis $\mathbf{\hat{e}}^{p}$, $\mathbf{%
\hat{e}}^{\bar{p}}$; so we shall drop the ${\hat{}}$ from all expressions.
It is convenient to define a real vector field $K$ on the K\"{a}hler base as
follows

\begin{equation}
K^{p}=if^{2}\mu ^{p},\qquad K^{\bar{p}}=-if^{2}(\mu ^{p})^{\ast }.
\end{equation}

In order to rewrite the constraints, we define a time co-ordinate $t$ so
that the Killing vector field associated with the Killing spinor $\epsilon
^{a}$ is
\begin{equation}
V={\frac{\partial }{\partial t}}
\end{equation}%
and set
\begin{equation}
\mathbf{e}^{0}=f^{2}(dt+\Omega )
\end{equation}%
where $\Omega $ is a 1-form defined on the K\"{a}hler base.

Then ({\ref{simp8}}) is equivalent to

\begin{equation}
\nabla _{p}K_{\bar{q}}={\frac{1}{\sqrt{2}f}}\partial _{t}\lambda \delta _{p%
\bar{q}}+\Omega _{p}\partial _{t}K_{\bar{q}}  \label{conf1}
\end{equation}%
where here $\nabla $ denotes the Levi-civita connection of the K\"{a}hler
base metric given in ({\ref{confmet}}). Also, ({\ref{simp11}}) can be
rewritten as

\begin{equation}
\nabla _{p}K_{q}={\frac{1}{\sqrt{2}f}}\partial _{t}\sigma ^{\ast }\epsilon
_{pq}+\Omega _{p}\partial _{t}K_{q}.  \label{conf2}
\end{equation}
It is also useful to define

\begin{equation}
Z=i_{K}J.
\end{equation}
It is then straightforward to show that

\begin{eqnarray}
\nabla _{p}Z_{\bar{q}} &=&{\frac{1}{\sqrt{2}f}}\partial _{t} \sigma
^{\ast}\delta _{p\bar{q}}+\Omega _{p}\partial _{t}Z_{\bar{q}},  \notag
\label{conf3} \\
\nabla _{p}Z_{q} &=&-{\frac{1}{\sqrt{2}f}}\partial _{t}\lambda \epsilon
_{pq}+\Omega _{p}\partial _{t}Z_{q}.
\end{eqnarray}
The commutator is given by

\begin{eqnarray}
\left[ K,Z\right] ^{p} &=&-i{\frac{\sqrt{2}}{f}}\left( K^{p}\partial _{t}(%
\func{Im}\sigma )+Z^{p}\partial _{t}(\func{Im}\lambda )\right) +(i_{K}\Omega
)\partial _{t}Z^{p}-(i_{Z}\Omega )\partial _{t}K^{p},  \notag
\label{eqn:commutx1} \\
\left[ K,Z\right] {}^{\bar{p}} &=&i{\frac{\sqrt{2}}{f}}\left( K^{\bar{p}%
}\partial _{t}(\func{Im}\sigma )+Z^{\bar{p}}\partial _{t}(\func{Im}\lambda
)\right) +(i_{K}\Omega )\partial _{t}Z^{\bar{p}}-(i_{Z}\Omega )\partial
_{t}K^{\bar{p}},
\end{eqnarray}%
Next, ({\ref{simp1}}) and ({\ref{simp3}}) are equivalent to

\begin{equation}
K^{p}\nabla _{p}X^{I}=-\sqrt{2} i \chi fV_{J}(X^{I}X^{J} -{\frac{3}{2}}%
Q^{IJ})\func{Im}\sigma ,  \label{nextsimp4}
\end{equation}

\begin{equation}
Z^{\bar{p}}\nabla _{\bar{p}}X^{I}=\sqrt{2}i \chi fV_{J}(X^{I}X^{J}-{\frac{3}{%
2}}Q^{IJ})\func{Im}\lambda .  \label{nextsimp5}
\end{equation}
These equations simply imply that

\begin{equation}
\mathcal{L}_{K}X^{I}=\mathcal{L}_{Z}X^{I}=0.  \label{scon}
\end{equation}%
In addition, ({\ref{simp4}}) and ({\ref{simp6}}) can be rewritten as

\begin{equation}
\partial _{t}\lambda =2\left( \sqrt{2}K^{p}\nabla _{p}f-i\chi f^{2}V_{I}X^{I}%
\func{Im}\sigma \right) ,  \label{nextsimp6}
\end{equation}%
and
\begin{equation}
\partial _{t}\sigma =2\left( \sqrt{2}Z^{\bar{p}}\nabla _{\bar{p}}f+i\chi
f^{2}V_{I}X^{I}\func{Im}\lambda \right) .  \label{nextsimp7}
\end{equation}%
In order to simplify the remainder of the equations, observe that for
indices $\mu ,\nu \neq 0$,

\begin{equation}
\omega _{0,\mu \nu }=-{\frac{1}{2}}f^{4}{\hat{(d\Omega )}}_{\mu \nu }
\label{spinconf1}
\end{equation}%
and

\begin{equation}
\omega _{p,qr}=-f{\hat{\omega}}_{p,qr}  \label{spinconf2}
\end{equation}
where on the LHS of ({\ref{spinconf1}}) and ({\ref{spinconf2}}), spatial
indices are taken with respect to the original five-dimensional basis,
whereas on the RHS, they are taken with respect to the conformally rescaled K%
\"{a}hler basis; and ${\hat{\omega}}$ denotes the spin connection of the K%
\"{a}hler base space. From henceforth, the hat will be dropped, and we will
work solely on the K\"{a}hler base space.

Then ({\ref{simp5}}) is equivalent to

\begin{equation}
\partial _{t}K_{\bar{q}}=\frac{1}{3}f^{6}K^{p}\left( {2}d\Omega _{p\bar{q}%
}-d\Omega _{m}{}^{m}\delta _{p\bar{q}}\right) +2\chi f^{2}V_{I}X^{I}Z_{\bar{q%
}}  \label{nextsimp9}
\end{equation}%
and using this, ({\ref{simp2}}) can be rewritten as

\begin{equation}
K^{p}F^{I}{}_{p\bar{q}}=f^{2}X^{I}K^{p}{d}\Omega _{p\bar{q}}+\frac{3\chi }{%
f^{2}}(X^{I}X^{J}-{\frac{1}{2}}Q^{IJ})V_{J}Z_{\bar{q}}-\frac{1}{f^{4}}%
\partial _{t}K_{\bar{q}}X^{I}.  \label{nextsimp8}
\end{equation}%
It is also useful to rewrite ({\ref{field}}) as

\begin{equation}
F^{I}{}_{pq}=f^{2}X^{I}{d}\Omega _{pq}+\frac{3\chi }{f^{2}}(X^{I}X^{J}-{%
\frac{1}{2}}Q^{IJ})V_{J}\epsilon _{pq}\ .  \label{nextsimp8a}
\end{equation}%
In addition, ({\ref{simp7}}), ({\ref{simp9}}), ({\ref{simp10}}) and ({\ref%
{simp12}}) can be rewritten as

\begin{equation}
\nabla _{p}(\func{Re}\frac{\lambda }{f})-\frac{1}{\sqrt{2}}(i_{K}d\Omega
)_{p}+\frac{1}{\sqrt{2}f^{6}}\partial _{t}K_{p}-\frac{\Omega _{p}}{f}%
\partial _{t}\func{Re}\lambda =0,  \label{nextsimp10}
\end{equation}

\begin{equation}
\nabla _{p}(\func{Re}\frac{\sigma }{f})-\frac{1}{\sqrt{2}}(i_{Z}d\Omega
)_{p}+\frac{1}{\sqrt{2}f^{6}}\partial _{t}Z_{p}-\frac{\Omega _{p}}{f}%
\partial _{t}\func{Re}\sigma =0,  \label{nextsimp11}
\end{equation}

\begin{eqnarray}
\nabla _{p}(\func{Im}\frac{\lambda }{f}) &=&{\frac{i}{2\sqrt{2}}}(d\Omega
_{mn}\epsilon ^{mn}+\frac{2\chi }{f^{4}}V_{I}X^{I})Z_{p}-i{\frac{\sqrt{2}}{6}%
}(K^{\bar{q}}d\Omega _{p\bar{q}}+K_{p}d\Omega _{q}{}^{q})  \notag \\
&&+\frac{1}{f}\omega _{p,mn}\epsilon ^{mn}\func{Im}\sigma +\frac{\Omega _{p}%
}{f}\partial _{t}\func{Im}\lambda ,  \label{nextsim12}
\end{eqnarray}

\begin{eqnarray}
\nabla _{p}(\func{Im}\frac{\sigma }{f}) &=&{\frac{i}{2\sqrt{2}}}
(d\Omega_{mn}\epsilon ^{mn}+\frac{2}{f^{4}}\chi V_{I}X^{I})K_{p} +{\frac{i%
\sqrt{2}}{6}}(Z^{\bar{q}}d\Omega _{p\bar{q}}+Z_{p}d\Omega _{q}{}^{q})  \notag
\\
&&-\frac{1}{f}\omega _{p,mn}\epsilon ^{mn}\func{Im}\lambda + \frac{\Omega
_{p}}{f}\partial _{t}\func{Im}\sigma .  \label{nextsim13}
\end{eqnarray}

\section{Half Supersymmetric Solutions}

Suppose that the solution preserves exactly four of the supersymmetries.
Then the four linearly independent Killing spinors are $\epsilon^{1},%
\epsilon^{2}$, $\eta^{1},\eta^{2}$ and

\begin{eqnarray}
(\epsilon ^{1})^{\prime } &=&\epsilon ^{2},\qquad (\epsilon ^{2})^{\prime
}=-\epsilon ^{1},  \notag \\
(\eta ^{1})^{\prime } &=&\eta ^{2},\qquad (\eta ^{2})^{\prime }=-\eta ^{1}.
\end{eqnarray}%
As all of the scalars, gauge field strengths and components of the spin
connection are $t$-independent, it follows that $\partial _{t}\eta
^{1},\partial _{t}\eta ^{2}$ is also a Killing spinor. As the solution is
exactly half-supersymmetric, it follows that there must be real constants $%
c_{1},c_{2},c_{3},c_{4}$ such that
\begin{equation}
\partial _{t}\eta ^{1}=c_{1}\eta ^{1}+c_{2}\eta ^{2}+c_{3}\epsilon
^{1}+c_{4}\epsilon ^{2}
\end{equation}%
or equivalently
\begin{eqnarray}
\partial _{t}\lambda &=&c_{1}\lambda -c_{2}\sigma ^{\ast }+c_{3}f,  \notag \\
\partial _{t}\sigma &=&c_{1}\sigma +c_{2}\lambda ^{\ast }+c_{4}f,  \notag \\
\partial _{t}\mu ^{p} &=&c_{1}\mu ^{p}-c_{2}\epsilon _{\bar{q}}{}^{p}(\mu
^{q})^{\ast }.  \label{tde}
\end{eqnarray}%
On substituting these constraints into ({\ref{simp5}}) we find that

\begin{equation}
\frac{1}{f^{2}}(c_{1}\mu_{\bar{q}}+c_{2}\epsilon_{\bar{q}\bar{n}}
(\mu^{n})^{\ast })+\frac{2}{3}\mu^{m}({2}\omega_{0,m\bar{q}}
-\omega_{0,p}{}^{p}\delta_{m{\bar{q}}})-2\chi V_{I}X^{I} \epsilon_{\bar{m}%
\bar{q}}(\mu^{m})^{\ast }=0.
\end{equation}
Contracting this expression with $(\mu^{q})^{\ast }$ we find that

\begin{equation}
(\mu^{q})^{\ast }\left( \frac{c_{1}}{f^{2}}\mu _{\bar{q}}+{\frac{2}{3}}\mu
^{m}({2}\omega_{0,m\bar{q}}-\omega_{0,p}{}^{p}\delta_{m{\bar{q}}})\right) =0.
\end{equation}
The real part of this expression implies that $c_{1}=0$.

Suppose now that $c_{2}\neq 0$. From ({\ref{tde}}) we find that
\begin{equation}
\partial_{t}\mu^{p}=-c_{2}\epsilon_{\bar{q}}{}^{p}(\mu^{q})^{\ast }
\end{equation}
and hence

\begin{equation}
\mu^{p}=\alpha^{p}\cos (c_{2}t)+\epsilon^{p}{}_{\bar{q}} (\alpha^{q})^{\ast
}\sin (c_{2}t)  \label{degencase1}
\end{equation}
where $\partial_{t}\alpha ^{p}=0$. Note that by making a redefinition of the
type

\begin{eqnarray}
\lambda &=&\lambda^{\prime }-{\frac{c_{4}}{c_{2}}}f,  \notag \\
\sigma &=&\sigma^{\prime }+{\frac{c_{3}}{c_{2}}}f,
\end{eqnarray}
we can without loss of generality set $c_{3}=c_{4}=0$ and drop the primes on
$\lambda $ and $\sigma $.

It will be convenient to split the solutions into three classes. For the
first class $c_{2}\neq 0$ and $c_{3}=c_{4}=0$, for the second $c_{2}=0$ but $%
c_{3}{}^{2}+c_{4}{}^{2}\neq 0$ and for the third $c_{2}=c_{3}=c_{4}=0$.

\subsection{Solutions with $c_2 \neq 0$, $c_3=c_4=0$}

For this class of solutions we have

\begin{eqnarray}
\partial_{t}\lambda &=&-c\sigma^{\ast },  \notag \\
\partial_{t}\sigma &=&c\lambda^{\ast },  \notag \\
\partial_{t}\mu ^{p} &=&c\epsilon^{p}{}_{\bar{q}}(\mu ^{q})^{\ast }
\end{eqnarray}
where $c=c_{2}\neq 0$. Here we have the conditions $\partial _{t}K=-cZ$ and $%
\partial _{t}Z=cK$.

To proceed with the analysis for these solutions, we define the 1-forms $%
\phi $, $\psi $ and $L$ on the K\"{a}hler base via

\begin{eqnarray}
L_{p} &=&\frac{1}{f}\left( \lambda Z_{p}-\sigma^{\ast }K_{p}\right) , \text{%
\ \ \ }L_{\bar{p}}=\frac{1}{f}\left( \lambda^{\ast }Z_{_{\bar{p}}}-\sigma
K_{_{\bar{p}}}\right)  \notag \\
\psi_{p} &=&\frac{1}{f}\left( \lambda^{\ast }Z_{p}-\sigma K_{p}\right) ,
\text{ \ \ \ }\psi _{\bar{p}}=\frac{1}{f}\left( \lambda Z_{\bar{p}}-
\sigma^{\ast }K_{\bar{p}}\right) ,  \notag \\
\phi_{p} &=&\frac{1}{f}\left( \lambda K_{p}+\sigma^{\ast }Z_{p}\right) ,
\text{ \ \ \ \ \ }\phi_{\bar{p}}=\frac{1}{f}\left( \lambda^{\ast } K_{\bar{p}%
}+\sigma Z_{\bar{p}}\right) .  \label{def}
\end{eqnarray}
The components of these 1-forms can be easily shown to be $t$-independent

\begin{equation}
\partial_{t}L_{p}=\partial_{t}\psi_{p}=\partial_{t}\phi_{p}=0.
\end{equation}
For convenience we set $\xi ^{2}=|\lambda |^{2}+|\sigma |^{2}$ and $%
z=(\lambda^{\ast })^{2}+\sigma^{2}.$

In order to evaluate various integrability constraints, it is useful to
compute the components of the covariant derivatives:

\begin{eqnarray}
\nabla_{p}L_{q} &=&-\frac{1}{\sqrt{2}}\left( d\Omega_{mn}\epsilon^{mn}+ {%
\frac{3}{f^{4}}}\chi V_{I}X^{I}\right) (K_{p}K_{q}+Z_{p}Z_{q}),  \notag \\
\nabla_{p}L_{\bar{q}} &=&{\frac{1}{\sqrt{2}}}d\Omega_{m}{}^{m}(Z_{p} K_{\bar{%
q}}-K_{p}Z_{\bar{q}})+{\frac{c\xi ^{2}}{\sqrt{2}f^{2}}} \delta _{p\bar{q}}
\notag \\
&&-{\frac{1}{\sqrt{2}}}\left( 3\frac{\chi V_{I}X^{I}}{f^{4}}+\frac{c}{f^{6}}
\right) (K_{p}K_{\bar{q}}+Z_{p}Z_{\bar{q}}),
\end{eqnarray}
and

\begin{eqnarray}
\nabla_{p}\psi _{q} &=&-{\frac{1}{\sqrt{2}}}\left( \frac{3\chi V_{I}X^{I}} {%
f^{4}}+\frac{c}{f^{6}}\right) (K_{p}K_{q}+Z_{p}Z_{q})+{\frac{1}{\sqrt{2}}}
d\Omega_{m}{}^{m}(Z_{p}K_{q}-K_{p}Z_{q})  \notag \\
&&-{\frac{\sqrt{2}ic}{f^{2}}}\func{Im}(\lambda \sigma )\epsilon_{pq},
\label{saf1} \\
\nabla _{p}\psi_{\bar{q}} &=&-\frac{1}{\sqrt{2}}\left(
d\Omega_{mn}\epsilon^{mn}+{\frac{3}{f^{4}}}\chi V_{I}X^{I}\right) (K_{p}K_{%
\bar{q}}+Z_{p}Z_{\bar{q}}) +{\frac{cz^{\ast }}{\sqrt{2}f^{2}}}\delta _{p\bar{%
q}},  \label{saf2}
\end{eqnarray}
and

\begin{eqnarray}
\nabla_{p}\phi_{q} &=&\frac{1}{\sqrt{2}}\left( d\Omega _{mn}\epsilon ^{mn}+ {%
\frac{3}{f^{4}}}\chi V_{I}X^{I}\right) (K_{p}Z_{q}-K_{q}Z_{p}) +{\frac{c}{%
\sqrt{2}f^{2}}}z^{\ast }\epsilon_{pq},  \label{saf3} \\
\nabla_{p}\phi_{\bar{q}} &=&-{\frac{1}{\sqrt{2}}} d\Omega_{m}{}^{m}(K_{p}K_{%
\bar{q}}+Z_{p}Z_{\bar{q}}) +{\frac{1}{\sqrt{2}}}(\frac{3\chi V_{I}X^{I}}{%
f^{4}}+\frac{c}{f^{6}}) (K_{p}Z_{\bar{q}}-K_{\bar{q}}Z_{p})  \notag \\
&&+{\frac{\sqrt{2}ic}{f^{2}}}\func{Im}(\lambda \sigma )\delta_{p\bar{q}}.
\label{saf4}
\end{eqnarray}
It immediately follows that $dL=0$ and $\phi $ defines a Killing vector on
the K\"{a}hler base space. In particular, setting $K^{2}=2K_{p}K^{p}$, $L$
is exact and satisfies

\begin{equation}
dK^{2}=\sqrt{2}cL.
\end{equation}
The first integrability condition we shall examine is obtained by
considering the constraints ({\ref{nextsimp4}}), ({\ref{nextsimp5}}), ({\ref%
{nextsimp6}}) and ({\ref{nextsimp7}}). These are equivalent to
\begin{equation}
d(\frac{X_{I}}{f^{2}})=\frac{\sqrt{2}}{K^{2}}\left( {\chi }V_{I}(\psi -L)-{c}
\frac{X_{I}}{f^{2}}L\right) .  \label{eqn:dfxcon1}
\end{equation}
Taking the exterior derivative of this equation, we obtain the constraint
\begin{equation}
d\psi =0.
\end{equation}
Hence, using (\ref{saf1}) and (\ref{saf2}) we obtain the constraints
\begin{equation}
d\Omega_{m}{}^{m}={\frac{4ic}{f^{2}K^{2}}}\func{Im}(\lambda \sigma ),
\end{equation}
and
\begin{equation}
{\frac{1}{2}}K^{2}(d\Omega _{\bar{m}\bar{n}} \epsilon^{\bar{m}\bar{n}%
}-d\Omega _{mn}\epsilon^{mn})-\frac{2ic}{f^{2}}\func{Im}z=0.
\end{equation}
Observe also that ({\ref{nextsimp9}}) is equivalent to

\begin{equation}
d\Omega_{p\bar{q}}={\frac{2ic}{f^{2}K^{2}}}\func{Im}(\lambda \sigma
)\delta_{p\bar{q}} -{\frac{3}{f^{4}K^{2}}}(2\chi V_{I}X^{I}+\frac{c}{f^{2}})
(K_{p}Z_{\bar{q}}-Z_{p}K_{\bar{q}}) \ .
\end{equation}
Using (\ref{saf3}) and (\ref{saf4}), we compute

\begin{equation}
d\phi =\sqrt{2}\left( 3\frac{\chi V_{I}X^{I}}{f^{4}}+\frac{c}{f^{6}}\right)
K\wedge Z+\frac{1}{\sqrt{2}}\left( K^{2}(d\Omega _{mn}\epsilon ^{mn} -\frac{c%
}{f^{6}})-2\frac{cz^{\ast }}{f^{2}}\right) J \ .
\end{equation}
To proceed, we impose the integrability condition, $d^{2}\phi =0$. We note
the following useful identities:

\begin{eqnarray}
\sqrt{2}d\left( \frac{3}{f^{4}}\chi V_{I}X^{I}+\frac{c}{f^{6}}\right) &=& {%
\frac{3V_{I}}{f^{2}K^{2}}}\left( -3\chi ^{2}(Q^{IJ}-2X^{I}X^{J})V_{J} +2c%
\frac{\chi }{f^{2}}X^{I}\right) (\psi -L)  \notag \\
&&-{\frac{6c}{f^{4}K^{2}}}(2\chi V_{I}X^{I}+\frac{c}{f^{2}})L
\end{eqnarray}
and

\begin{eqnarray}
L\wedge K\wedge Z &=&\frac{K^{2}}{2}(L-\psi )\wedge J,  \notag \\
\psi \wedge K\wedge Z &=&-\frac{K^{2}}{2}(L-\psi )\wedge J,  \notag \\
d(K\wedge Z) &=&-{\frac{c}{\sqrt{2}}}(3\psi -L)\wedge J
\end{eqnarray}
from whence we obtain

\begin{equation}
\epsilon^{pq}d\left( (\frac{3\chi V_{I}X^{I}}{f^{4}}+\frac{c}{f^{6}})
K\wedge Z\right)_{pq\bar{\ell}}=-d\left( K^{2}(\frac{6\chi V_{I}X^{I}} {f^{4}%
}+\frac{c}{f^{6}})\right)_{\bar{\ell}}.
\end{equation}
Using this expression, the constraint $d^{2}\phi =0$ implies that

\begin{equation}
d\Omega_{mn}\epsilon ^{mn}={\frac{\sqrt{2}\theta }{K^{2}}}+{\frac{2cz^{\ast}%
} {f^{2}K^{2}}}-\frac{6\chi V_{I}X^{I}}{f^{4}}
\end{equation}
for real constant $\theta $.

Next we consider the integrability condition $d^2 \Omega=0$. It is
straightforward to show that $d \Omega$ satisfies

\begin{equation}
d\Omega =-\sqrt{2}d\left( {\frac{\phi }{K^{2}}}\right) +\frac{c}{f^{2}}
\left( {\frac{1}{2f^{4}}}-{\frac{\xi ^{2}}{K^{2}}}\right) \left( J- {\frac{2%
}{K^{2}}}K\wedge Z\right) +{\frac{\theta }{\sqrt{2}K^{2}}}J.
\label{domconstr1}
\end{equation}
Hence the integrability condition $d^{2}\Omega =0$ implies that
\begin{equation}
d\left( (\frac{c}{2f^{6}}-{\frac{c\xi ^{2}}{f^{2}K^{2}}})(J-{\frac{2}{K^{2}}}
K\wedge Z)+{\frac{\theta }{\sqrt{2}K^{2}}}J\right) =0.  \label{domintega}
\end{equation}
We observe that

\begin{equation}
\nabla_{p}(\frac{1}{f^{6}}-{\frac{2\xi ^{2}}{f^{2}K^{2}}})= {\frac{2}{%
(K^{2})^{2}}}(\theta +\frac{\sqrt{2}}{f^{2}}cz^{\ast })\psi_{p} +\frac{2%
\sqrt{2}c}{f^{2}K^{2}}({\frac{\xi ^{2}}{K^{2}}}-\frac{{1}}{f^{4}})L_{p}+ {%
\frac{8\sqrt{2i}c}{f^{2}(K^{2})^{2}}}\func{Im}(\lambda \sigma )\phi_{p} \ .
\end{equation}
It is then straightforward but tedious to show that ({\ref{domintega}})
implies

\begin{equation}
\func{Im}\lambda \sigma =0.  \label{domegarealcon}
\end{equation}
Using this condition the above relations simplify and we obtain

\begin{eqnarray}
d\Omega_{m}{}^{m} &=&{0,}  \notag \\
\nabla_{p}L_{q} &=&-\left( {\frac{\theta }{K^{2}}}+{\frac{\sqrt{2}c} {%
f^{2}K^{2}}}z^{\ast }-\frac{3}{\sqrt{2}f^{4}}\chi V_{I}X^{I}\right)
(K_{p}K_{q}+Z_{p}Z_{q}),  \notag \\
\nabla_{p}L_{\bar{q}} &=&-{\frac{1}{\sqrt{2}}}(\frac{3\chi }{f^{4}}
V_{I}X^{I}+\frac{c}{f^{6}})(K_{p}K_{\bar{q}}+Z_{p}Z_{\bar{q}})+ {\frac{c}{%
\sqrt{2}f^{2}}}\xi ^{2}\delta_{p\bar{q}},  \notag \\
\nabla _{p}\psi _{q} &=&-{\frac{1}{\sqrt{2}}}(\frac{3\chi }{f^{4}}%
V_{I}X^{I}+ \frac{c}{f^{6}})(K_{p}K_{q}+Z_{p}Z_{q}),  \notag \\
\nabla_{p}\psi_{\bar{q}} &=&-\left( {\frac{\theta }{K^{2}}}+ {\frac{\sqrt{2}c%
}{f^{2}K^{2}}}z^{\ast }-\frac{3}{\sqrt{2}f^{4}}\chi V_{I}X^{I}\right)
(K_{p}K_{\bar{q}}+Z_{p}Z_{\bar{q}})+{\frac{cz^{\ast }}{\sqrt{2}f^{2}}}\delta
_{p\bar{q}},  \notag \\
\nabla_{p}\phi_{q} &=&\left( {\frac{\theta }{K^{2}}}+{\frac{\sqrt{2}c} {%
f^{2}K^{2}}}z^{\ast }-\frac{3}{\sqrt{2}f^{4}}\chi V_{I}X^{I}\right)
(K_{p}Z_{q}-K_{q}Z_{p})+{\frac{cz^{\ast }}{\sqrt{2}f^{2}}}\epsilon _{pq},
\notag \\
\nabla_{p}\phi_{\bar{q}} &=&{\frac{1}{\sqrt{2}}}(\frac{3\chi }{f^{4}}
V_{I}X^{I}+\frac{c}{f^{6}})(K_{p}Z_{\bar{q}}-K_{\bar{q}}Z_{p}).
\end{eqnarray}

\bigskip

The components of $d\Omega $ are therefore given by

\begin{eqnarray}
d\Omega_{p\bar{q}} &=&-{\frac{1}{f^{4}K^{2}}}(6\chi V_{I}X^{I} +\frac{3c}{%
f^{2}})(K_{p}Z_{\bar{q}}-Z_{p}K_{\bar{q}})  \notag \\
d\Omega_{mn}\epsilon ^{mn} &=&{\frac{2c}{f^{2}K^{2}}}z^{\ast }-\frac{6\chi }{%
f^{4}}V_{I}X^{I}+{\frac{\sqrt{2}\theta }{K^{2}}} \ .
\end{eqnarray}
Using the expressions for $d\Omega $ which we have obtained, we next examine
({\ref{nextsimp10}}) and ({\ref{nextsimp11}}). These may be rewritten as

\begin{equation}
\nabla_{p}(\frac{\func{Re}\lambda }{f})=-\frac{1}{2}\left( {\frac{\theta }{%
K^{2}}}+\frac{c}{\sqrt{2}f^{6}}+{\frac{\sqrt{2}c}{f^{2}K^{2}}}
z^{\ast}\right) Z_{p}-c\Omega _{p}\func{Re}\left( \frac{\sigma }{f}\right)
\label{lam}
\end{equation}
and
\begin{equation}
\nabla_{p}(\frac{\func{Re}\sigma }{f})=-\frac{1}{2}\left( {\frac{\theta }{%
K^{2}}}+\frac{c}{\sqrt{2}f^{6}}+{\frac{\sqrt{2}c}{f^{2}K^{2}}}
z^{\ast}\right) K_{p}+c\Omega _{p}\func{Re}\left( \frac{\lambda }{f}\right) .
\label{sig}
\end{equation}
We note the useful identities:

\begin{eqnarray}
\nabla_{p}\Sigma &=&{\frac{3c\chi }{f^{4}K^{2}}}V_{I}X^{I} (\psi_{p}+L_{p})-{%
\frac{3\sqrt{2}c}{K^{2}}}\Sigma L_{p} \\
\nabla _{p}\Sigma ^{\ast } &=&-{\frac{c}{f^{4}K^{2}}}(3\chi V_{I}X^{I} +%
\frac{2c}{f^{2}})(\psi _{p}+L_{p})-{\frac{\sqrt{2}c}{K^{2}}}\Sigma ^{\ast
}L_{p}
\end{eqnarray}
where we have set $\Sigma ={\frac{\theta }{K^{2}}}+\frac{c}{\sqrt{2}f^{6}} +{%
\frac{\sqrt{2}c}{f^{2}K^{2}}}z^{\ast }.$ Then from the integrability
condition $\epsilon^{mn}\nabla _{m}\nabla _{n} (\func{Re}\frac{\lambda }{f}%
)=0,$ we find the constraint

\begin{equation}
c\sigma ^{\ast }\left( {\frac{\theta }{K^{2}}}+{\frac{c}{\sqrt{2}f^{6}}}- {%
\frac{\sqrt{2c}}{f^{2}K^{2}}}\xi ^{2}\right) +\left( {3}\sqrt{2}\frac{c\chi
V_{I}X^{I}}{f^{4}}-2{\frac{c\theta }{K^{2}}}\right) \func{Re}\sigma + {\frac{%
2ic}{fK^{2}}}Z^{p}\omega _{p,mn}\epsilon^{mn}\func{Im}z=0.  \label{rr2}
\end{equation}
Note that
\begin{equation*}
\partial_{t}\func{Im}z=\partial _{t}\xi ^{2}=0,
\end{equation*}
then upon differentiating (\ref{rr2}) with respect to $t$ gives

\begin{equation}
c\lambda \left( {\frac{\theta }{K^{2}}}+{\frac{c}{\sqrt{2}f^{6}}}- {\frac{%
\sqrt{2c}}{f^{2}K^{2}}}\xi ^{2}\right) +\left( {3}\sqrt{2}\frac{c\chi
V_{I}X^{I}}{f^{4}}-2{\frac{c\theta }{K^{2}}}\right) \func{Re}\lambda +{\frac{%
2ic}{fK^{2}}}K^{p}\omega _{p,mn}\epsilon ^{mn}\func{Im}z=0.  \label{rrt}
\end{equation}
It turns out that the constraints (\ref{rr2}) and (\ref{rrt}) are also
sufficient to ensure that
\begin{equation}
\nabla _{\lbrack p}\nabla _{\bar{q}]}(\func{Re}\frac{\lambda }{f})=\nabla
_{\lbrack p}\nabla _{\bar{q}]}(\func{Re}\frac{\sigma }{f})=0.
\end{equation}

\bigskip Next, note that the constraints ({\ref{nextsimp8}}) and ({\ref%
{nextsimp8a}}) can be used to write the gauge field strengths $F^{I}$ as

\begin{eqnarray}
F^{I} &=&d\left( f^{2}X^{I}(dt+\Omega )\right) +\frac{6\chi }{f^{2}}
V_{J}(X^{I}X^{J}-{\frac{1}{2}}Q^{IJ})\left( {\frac{1}{K^{2}}}K\wedge
Z-J\right)  \notag  \label{Fconst1} \\
&+&\frac{c}{f^{4}}X^{I}\left( {\frac{2}{K^{2}}}K\wedge Z-J\right) .
\end{eqnarray}
and note that as $(L-\psi )\wedge \left( {\frac{1}{K^{2}}}K\wedge Z-J\right)
=0$ it follows that
\begin{equation*}
d(X^{I}X^{J}-{\frac{1}{2}}Q^{IJ})V_{J}\wedge \left( {\frac{1}{K^{2}}}K\wedge
Z-J\right) =0 \ .
\end{equation*}
It is then straightforward to show that the Bianchi identity $dF^{I}=0$
follows automatically from the constraints we have obtained. To proceed
further, it is useful to consider the cases for which $\func{Im}z=0$ and $%
\func{Im}z\neq 0$ separately. Observe that $\func{Im}z=0$ implies that $%
\lambda $ and $\sigma $ are either both real or both imaginary.

\subsubsection{Solutions with $\func{Im}z\neq 0$}

In order to introduce a local co-ordinate system for solutions with $\func{Im%
}z\neq 0$, recall that $\phi $ is a Killing vector on the K\"{a}hler base
space. Furthermore, as $\psi =i_{\phi }J$, the closure of $\psi $ implies
that $\phi $ preserves the K\"{a}hler form;
\begin{equation}
\mathcal{L}_{\phi }J=0.
\end{equation}
It is also straightforward to show that
\begin{equation}
\mathcal{L}_{\phi }X_{I}=\mathcal{L}_{\phi }f=\mathcal{L}_{\phi }d\Omega =
\mathcal{L}_{\phi }F^{I}=0.
\end{equation}
Hence it follows that $\phi $ defines a symmetry of the full five
dimensional solution. As $\func{Im}z\neq 0$, ({\ref{rr2}}) and ({\ref{rrt}})
can be inverted to obtain

\begin{equation}
\omega_{p,mn}\epsilon^{mn}={\frac{i}{\sqrt{2}\func{Im}z}} \left( \frac{c}{%
f^{4}}-{\frac{2c\xi ^{2}}{K^{2}}}+3\chi V_{I}X^{I} \left( \frac{1}{f^{2}}+%
\frac{z}{f^{2}\xi ^{2}}\right) -\frac{\sqrt{2}z}{\xi ^{2}}{\frac{\theta f^{2}%
}{K^{2}}}\right) \phi _{p}.  \label{emb}
\end{equation}

It is convenient to define the real 1-forms ${\hat{L}}$ and ${\hat{\phi}}$
by
\begin{eqnarray}
{\hat{L}}_{p} &=&iL_{p},\qquad {\hat{L}}_{\bar{p}}=-iL_{\bar{p}},  \notag \\
{\hat{\phi}}_{p} &=&i\phi _{p},\qquad {\hat{\phi}}_{\bar{p}}=-i\phi _{\bar{p}%
}.
\end{eqnarray}
It is then straightforward, but tedious, to show that

\begin{equation}
d\left( {\frac{f^{2}}{(K^{2})^{2}\func{Im}z}}{\hat{L}}\right) =0,
\label{hyperaux1}
\end{equation}
and also that

\begin{equation}
\lbrack \phi ,L]=[\phi ,{\hat{L}}]=[\phi ,{\hat{\phi}}]=0.
\end{equation}
In addition, we find that

\begin{equation}
d\phi =\left( \frac{3\chi V_{I}X^{I}}{\sqrt{2}f^{2}\xi ^{2}}+\frac{c} {\sqrt{%
2}f^{4}\xi ^{2}}\right) (\phi \wedge L+{\hat{\phi}}\wedge {\hat{L}})+
(\theta -{\frac{3}{\sqrt{2}f^{4}}}\chi K^{2}V_{I}X^{I})J.
\label{dphidecomp1}
\end{equation}
Hence we define the following orthonormal basis on the K\"{a}hler base space

\begin{equation}
\mathbf{e}^{1}=\frac{\phi }{H},\quad \mathbf{e}^{2}=\frac{L}{H},\quad
\mathbf{e}^{3}=\frac{{\hat{\phi}}}{H},\quad \mathbf{e}^{4}=\frac{{\hat{L}}}{H%
}
\end{equation}%
where
\begin{equation}
H^{2}=\frac{K^{2}\xi ^{2}}{f^{2}}.
\end{equation}%
As $\phi $, ${\hat{\phi}}$ are commuting vector fields, we can choose
co-ordinates $\tau $, $\eta $ such that
\begin{equation}
\phi ={\frac{\partial }{\partial \tau }},\quad {\hat{\phi}}={\frac{\partial
}{\partial \eta }.}
\end{equation}%
Then, defining
\begin{equation}
v={\frac{K^{2}}{\sqrt{2}c},}
\end{equation}%
we have
\begin{equation}
L=dv
\end{equation}%
and from ({\ref{hyperaux1}}) we see that there must be a function $u$ such
that
\begin{equation}
{\hat{L}}={\frac{\func{Im}z}{\sqrt{2}f^{2}c}}(K^{2})^{2}du.
\end{equation}%
As $\phi $, ${\hat{\phi}}$, $L$, ${\hat{L}}$ are orthogonal, it follows that
$(\tau ,\eta ,u,v)$ form a local co-ordinate system on the base space. One
can then write
\begin{eqnarray}
\phi &=&H^{2}(d\tau +\alpha _{1}du+\alpha _{2}dv),\qquad  \notag \\
{\hat{\phi}} &=&H^{2}(d\eta +\beta _{1}du+\beta _{2}dv).
\end{eqnarray}%
As $\phi $ is a Killing vector, the functions $H$, $f^{-2}(K^{2})^{2}\func{Im%
}z$, $\alpha _{1}$, $\alpha _{2}$, $\beta _{1}$ and $\beta _{2}$ do not
depend on $\tau $ (or $t$). Furthermore,
\begin{equation*}
\mathcal{L}_{\hat{\phi}}f=\mathcal{L}_{_{\hat{\phi}}}X_{I}=\mathcal{L}_{_{%
\hat{\phi}}}\left( \frac{(K^{2})^{2}}{f^{2}}\func{Im}z\right) =0.
\end{equation*}%
Therefore $f$, $\ (K^{2})^{2}\func{Im}z$ and $X_{I}$ are functions of $u$
and $v$ only. However, there is a non-trivial $\eta $ -dependence in $\alpha
_{1},\alpha _{2},\beta _{2},\beta _{2}$; ${\hat{\phi}}$ is not a Killing
vector.

It is also useful to observe that the identity $\func{Im}\lambda \sigma =0$
implies

\begin{equation}
(\func{Re}z)^{2}+(\func{Im}z)^{2}=\xi ^{4},
\end{equation}
and hence we shall set

\begin{eqnarray}
\cos Y &=&{\frac{\func{Re}z}{\xi ^{2}},}  \notag \\
\sin Y &=&{\frac{\func{Im}z}{\xi ^{2}}}.
\end{eqnarray}
Here $Y$ is a real function which satisfies

\begin{equation}
\mathcal{L}_{\phi }Y=\mathcal{L}_{\hat{\phi}}Y=0
\end{equation}
which implies $Y=Y(u,v)$. With these conventions, it is straightforward to
compute

\begin{eqnarray}
{\frac{\partial H^{2}}{\partial u}} &=&H^{2}v\sin ^{2}Y\left( 3\frac{\chi
cvV_{I}X^{I}}{f^{4}}-\theta \right) ,  \notag \\
{\frac{\partial H^{2}}{\partial v}} &=&-\frac{cv}{f^{4}}\left( 3\chi
V_{I}X^{I}+\frac{c}{f^{2}}\right) +\cos Y(3\frac{\chi cvV_{I}X^{I}}{f^{4}}
-\theta ),  \label{Hcons2}
\end{eqnarray}
and

\begin{eqnarray}
{\frac{\partial Y}{\partial u}} &=&\sin Y\left( -H^{2}+3\frac{\chi
cv^{2}V_{I}X^{I}}{f^{4}}+\frac{c^{2}v^{2}}{f^{6}}\right) +\frac{v}{2}\sin
2Y\left( 3\frac{\chi cvV_{I}X^{I}}{f^{4}}-\theta \right) ,  \notag \\
{\frac{\partial Y}{\partial v}} &=&-\frac{1}{H^{2}}\sin Y\left( 3\frac{\chi
cvV_{I}X^{I}}{f^{4}}-\theta \right) .  \label{Ycons2}
\end{eqnarray}
Also note that ({\ref{eqn:dfxcon1}}) can be rewritten as

\begin{eqnarray}
\frac{\partial }{\partial u}(\frac{X_{I}}{f^{2}}) &=& \frac{{\chi }%
H^{2}V_{I}\sin ^{2}Y}{c},  \notag \\
{\frac{\partial }{\partial v}}(\frac{X_{I}}{f^{2}}) &=&\frac{1}{v}\left(
\frac{{\chi }V_{I}(\cos Y-1)}{c}-\frac{X_{I}}{f^{2}}\right) .
\end{eqnarray}
If $\theta \neq 0,$ then this constraint can be integrated up to give

\begin{equation}
X_{I}=f^{2}\left( {\frac{q_{I}}{v}+\frac{\chi }{c}}\left( \frac{c^{2}v}{%
f^{6}\theta }-{\frac{H^{2}}{\theta v}}-1\right) V_{I}\right) .
\end{equation}
for constant $q_{I}$.

To proceed, consider the equation ({\ref{dphidecomp1}}), which can be
rewritten as

\begin{equation}
d\phi =\frac{cv}{f^{6}}\left( 3\chi f^{2}V_{I}X^{I}+c\right) (\mathbf{e}%
^{1}\wedge \mathbf{e}^{2}+\mathbf{e}^{3}\wedge \mathbf{e}^{4})-\left( 3\frac{%
\chi cvV_{I}X^{I}}{f^{4}}-\theta \right) J.  \label{dphidecomp2}
\end{equation}%
Taking the self-dual projection of ({\ref{dphidecomp2}}) yields the
constraints

\begin{eqnarray}
{\frac{\partial \alpha _{1}}{\partial \eta }} &=&\frac{1}{2}v\sin 2Y\left( 3%
\frac{\chi cvV_{I}X^{I}}{f^{4}}-\theta \right) +\frac{cv^{2}\sin Y}{f^{6}}
\left( 3\chi f^{2}V_{I}X^{I}+c\right) ,  \notag \\
{\frac{\partial \alpha _{2}}{\partial \eta }} &=&-\frac{\sin Y}{H^{2}}\left(
3\frac{\chi cvV_{I}X^{I}}{f^{4}}-\theta \right)  \label{alphaco1}
\end{eqnarray}
together with

\begin{equation}
{\frac{\partial \alpha _{2}}{\partial u}}-{\frac{\partial \alpha _{1}} {%
\partial v}}+\beta _{2}{\frac{\partial \alpha _{1}}{\partial \eta }}-\beta
_{1}{\frac{\partial \alpha _{2}}{\partial \eta }}=0.  \label{alphaco2}
\end{equation}
Using these constraints, the anti-self-dual projection of ({\ref{dphidecomp2}%
}) fixes $J$ to be given by

\begin{equation}
J=\cos Y(\mathbf{e}^{1}\wedge \mathbf{e}^{2}-\mathbf{e}^{3}\wedge \mathbf{e}%
^{4})+\sin Y(\mathbf{e}^{1}\wedge \mathbf{e}^{4}-\mathbf{e}^{2}\wedge
\mathbf{e}^{3}).
\end{equation}%
Imposing the covariant constancy condition $\nabla J=0$ imposes two
additional constraints:

\begin{equation}
{\frac{\partial \beta _{2}}{\partial u}}-{\frac{\partial \beta _{1}} {%
\partial v}}+\beta _{2}{\frac{\partial \beta _{1}}{\partial \eta }}-\beta
_{1}{\frac{\partial \beta _{2}}{\partial \eta }}=0,  \label{betacon1}
\end{equation}
and

\begin{equation}
\sin Y{\frac{\partial \beta _{1}}{\partial \eta }}+\frac{1}{2}H^{2}v\sin 2Y {%
\frac{\partial \beta _{2}}{\partial \eta }}=-\frac{cv^{2}}{2f^{6}}\sin
2Y\left( 3\chi f^{2}V_{I}X^{I}+c\right) .  \label{betader1}
\end{equation}

Finally, we compare the spin connection components $\omega_{p,mn}%
\epsilon^{mn}$ computed in this basis with the expression given in ({\ref%
{emb}}), noting that $\omega_{p,mn}\epsilon^{mn}=-{\frac{1}{2}}\omega_{p,\mu
\nu}J^{\mu \nu}$.

This implies that

\begin{eqnarray}
-\cos Y{\frac{\partial \beta _{1}}{\partial \eta }}+H^{2}v\sin ^{2}Y {\frac{%
\partial \beta _{2}}{\partial \eta }} &=&\frac{cv^{2}}{f^{6}}\cos
^{2}Y\left( 3\chi f^{2}V_{I}X^{I}+c\right)  \notag \\
&&+v\cos Y\left( \frac{3\chi cvV_{I}X^{I}}{f^{4}}-\theta \right) -H^{2}.
\label{betade2}
\end{eqnarray}
Then from ({\ref{betader1}}) and ({\ref{betade2}}) we obtain

\begin{eqnarray}
{\frac{\partial \beta _{2}}{\partial \eta }} &=&\frac{1}{H^{2}}\cos Y\left(
\frac{3\chi cvV_{I}X^{I}}{f^{4}}-\theta \right) -{\frac{1}{v},}  \notag \\
{\frac{\partial \beta _{1}}{\partial \eta }} &=&-v\cos Y\left( H^{2} {\frac{%
\partial \beta _{2}}{\partial \eta }+}\frac{cv}{f^{6}}\left( 3\chi
f^{2}V_{I}X^{I}+c\right) \right) .  \label{ambu}
\end{eqnarray}
Then ({\ref{betacon1}}) and ({\ref{ambu}}) can be integrated up to give

\begin{equation}
\beta _{1}=-\eta {\cot Y\frac{\partial Y}{\partial u}},\quad \beta
_{2}=-\eta \left( {\cot Y\frac{\partial Y}{\partial v}}+{\frac{1}{v}}\right)
\end{equation}
and ({\ref{alphaco1}}) and ({\ref{alphaco2}}) then imply

\begin{equation}
\alpha _{1}=\eta \left( {\frac{\partial Y}{\partial u}}+H^{2}\sin Y\right)
,\quad \alpha _{2}=\eta {\frac{\partial Y}{\partial v}.}
\end{equation}

Hence, in these co-ordinates, the orthonormal basis of the K\"ahler base
space is

\begin{eqnarray}
\mathbf{e}^{1} &=&H\left( d\tau +\eta \left( {\frac{\partial Y}{\partial u}}%
+H^{2}\sin Y\right) du+\eta {\frac{\partial Y}{\partial v}}dv\right) ,
\notag \\
\mathbf{e}^{2} &=&\frac{1}{H}dv,  \notag \\
\mathbf{e}^{3} &=&H\left( d\eta -\eta {\cot Y}{\frac{\partial Y}{\partial u}}%
du-\eta \left( {\cot Y}{\frac{\partial Y}{\partial v}}+{\frac{1}{v}}\right)
dv\right)  \notag \\
\mathbf{e}^{4} &=&Hv\sin Ydu
\end{eqnarray}%
and if $\theta \neq 0$, then $J$ can be written as
\begin{equation*}
J=d\left( (\frac{H^{2}}{\theta }-\frac{c^{2}v^{2}}{\theta f^{6}})d\tau +\eta
\sin Ydv-\frac{1}{2}\eta H^{2}v\sin 2Ydu\right) .
\end{equation*}

By considering ({\ref{domconstr1}}), $\Omega$ is fixed (up to a total
derivative) by

\begin{eqnarray}
\Omega &=&-{\frac{1}{2cv}}(H^{2}+\frac{c^{2}v^{2}}{f^{6}})d\tau + {\frac{%
\eta }{cv}}({\frac{1}{2}}\theta \sin Y-H^{2}{\frac{\partial Y}{\partial v}}
)dv  \notag \\
&&-\eta \left( {\frac{H^{2}}{cv}}({\frac{\partial Y}{\partial u}}+H^{2}\sin
Y)+{\frac{1}{2}}\sin Y(\frac{{\theta }H^{2}}{c}\cos Y+\frac{c}{f^{6}}H^{2}v-
{\frac{1}{cv}}H^{4})\right) du  \notag \\
&&
\end{eqnarray}

and by considering ({\ref{Fconst1}}) we find the gauge field strengths are
given by

\begin{eqnarray}
F^{I} &=&d\big[f^{2}X^{I}(dt+\Omega )+\frac{cvX^{I}}{f^{4}}\left( d\tau
+\eta H^{2}\sin Ydu\right)  \notag \\
&&-\frac{3\chi \eta }{f^{2}}\sin Y(X^{I}X^{J}-{\frac{1}{2}}
Q^{IJ})V_{J}(-H^{2}v(1+\cos Y)du+dv))\big]
\end{eqnarray}

\subsubsection{Solutions with $\func{Re}\protect\lambda =\func{Re}\protect%
\sigma =0$, $\protect\lambda \neq 0,\protect\sigma \neq 0$}

If $\lambda $ and $\sigma $ are imaginary but non-vanishing, then from (\ref%
{def}) we obtain $\psi =-L$, and the constraints on $\lambda $ and $\sigma $
as given in (\ref{lam}) and (\ref{sig}) imply that

\begin{equation}
{\frac{\theta }{K^{2}}}+{\frac{c}{\sqrt{2}f^{6}}}={\frac{\sqrt{2}c} {%
f^{2}K^{2}}}\xi ^{2}.
\end{equation}
Moreover, ({\ref{eqn:dfxcon1}}) can be integrated up to give

\begin{equation}
X_{I}=f^{2}\left( -{\frac{2\chi }{c}}V_{I}+{\frac{\rho _{I}}{K^{2}}}\right)
\end{equation}
for constants $\rho _{I}$. The K\"{a}hler form can be expressed by

\begin{equation}
J=d\left( {\frac{\phi f^{6}}{\theta f^{6}+{\frac{c}{\sqrt{2}}}K^{2}}}\right)
\ .
\end{equation}

It can also be demonstrated that $\Omega $ is given in these cases by

\begin{equation}
\Omega =\left( -{\frac{c}{\theta f^{6}+{\frac{c}{\sqrt{2}}}K^{2}}}\phi
\right) +dg_{2}
\end{equation}
for some real function $g_{2}$ with

\begin{equation}
F^{I}=d\left( f^{2}X^{I}(dt+\Omega )+{\frac{cf^{2}X^{I}}{\theta f^{6}+ {%
\frac{c}{\sqrt{2}}}K^{2}}}\phi \right)
\end{equation}
and hence

\begin{equation}
\omega_{p,mn}\epsilon^{mn}=-{\frac{3c\chi f^{2}V_{I}X^{I}}{\theta f^{6}+ {%
\frac{c}{\sqrt{2}}}K^{2}}}\phi _{p}+\partial _{p}g_{1}
\end{equation}
for some real function $g_{1}$ where $g_{1},g_{2}$ satisfy

\begin{equation}
\partial_{p}\arctan \left( {\frac{\lambda }{\sigma }}\right)
=\partial_{p}(g_{1}+cg_{2}) \ .
\end{equation}
Note that $\arctan \left( {\frac{\lambda }{\sigma }}\right) =ct+H$ with $%
\partial _{t}H=0$. Without loss of generality, we can work in a gauge for
which $g_{2}={\frac{H}{c}}$ and $g_{1}=0$.

It is then straightforward to prove the following identities:

\begin{eqnarray}
d\left( {\frac{\phi }{K^{2}(c\theta +{\frac{c^{2}}{\sqrt{2}f^{6}}}K^{2})}}%
\right) &=&-{\frac{\sqrt{2}c}{(K^{2})^{2}(c\theta +{\frac{c^{2}}{\sqrt{2}f^{%
\emph{6}}}}K^{2})}}\hat{L}\wedge \hat{\phi},  \notag \\
d\left( {\frac{\hat{L}}{K^{2}\sqrt{c\theta +{\frac{c^{2}}{\sqrt{2}f^{6}}}%
K^{2}}}}\right) &=&{\frac{\sqrt{2}c^{2}\theta }{(K^{2})^{2}(c\theta +{\frac{%
c^{2}}{\sqrt{2}f^{6}}}K^{2})^{\frac{3}{2}}}}\phi \wedge \hat{\phi},  \notag
\\
d\left( {\frac{\hat{\phi}}{K^{2}\sqrt{c\theta +{\frac{c^{2}}{\sqrt{2}f^{6}}}%
K^{2}}}}\right) &=&{\frac{\sqrt{2}c^{2}\theta }{(K^{2})^{2}(c\theta +{\frac{%
c^{2}}{\sqrt{2}f^{6}}}K^{2})^{\frac{3}{2}}}}\hat{L}\wedge \phi .
\end{eqnarray}

There are then three cases to consider.

\begin{itemize}
\item[i)] If $c \theta>0$ then define
\end{itemize}

\begin{eqnarray}
\sigma ^{1} &=&{\frac{\sqrt{2}c^{2}\theta }{K^{2}(c\theta +{\frac{c^{2}}{%
\sqrt{2}f^{6}}}K^{2})}}\phi ,  \notag \\
\sigma ^{2} &=&{\frac{\sqrt{2c^{3}\theta }}{K^{2}\sqrt{c\theta +{\frac{c^{2}%
}{\sqrt{2}f^{6}}}K^{2}}}}\hat{L},  \notag \\
\sigma ^{3} &=&{\frac{\sqrt{2c^{3}\theta }}{K^{2}\sqrt{c\theta +{\frac{c^{2}%
}{\sqrt{2}f^{6}}}K^{2}}}}\hat{\phi}.
\end{eqnarray}%
It is then straightforward to show that
\begin{eqnarray}
d\sigma ^{i} &=&-{\frac{1}{2}}\epsilon _{ijk}\sigma ^{j}\wedge \sigma ^{k},%
\text{ \ }  \notag \\
\mathcal{L}_{L}\sigma ^{i} &=&0.
\end{eqnarray}

\begin{itemize}
\item[ii)] If $\theta =0$ then define
\end{itemize}

\begin{eqnarray}
\sigma ^{1} &=&-{\frac{f^6}{c^{3}(K^{2})^{2}}}\phi ,  \notag \\
\sigma ^{2} &=&{\frac{f^3}{K^{2}\sqrt{{\frac{c^{2}}{\sqrt{2}}}K^{2}}}} \hat{L%
},  \notag \\
\sigma ^{3} &=&{\frac{f^3}{K^{2}\sqrt{{\frac{c^{2}}{\sqrt{2}}}K^{2}}}} \hat{%
\phi},
\end{eqnarray}

where

\bigskip

\begin{eqnarray}
d\sigma ^{1} &=&\sigma ^{2}\wedge \sigma ^{3},  \notag \\
d\sigma ^{2} &=&d\sigma ^{3}=0,  \notag \\
\mathcal{L}_{L}\sigma ^{i} &=&0.
\end{eqnarray}

\begin{itemize}
\item[iii)] If $c\theta <0$ then define
\end{itemize}

\begin{eqnarray}
\sigma ^{1} &=&{\frac{\sqrt{2}c^{2}\theta }{K^{2}(c\theta +{\frac{c^{2}}{%
\sqrt{2}f^{6}}}K^{2})}}\phi ,  \notag \\
\sigma ^{2} &=&{\frac{\sqrt{-2c^{3}\theta }}{K^{2}\sqrt{c\theta +{\frac{c^{2}%
}{\sqrt{2}f^{6}}}K^{2}}}}\hat{L},  \notag \\
\sigma ^{3} &=&{\frac{\sqrt{-2c^{3}\theta }}{K^{2}\sqrt{c\theta +{\frac{c^{2}%
}{\sqrt{2}f^{6}}}K^{2}}}}\hat{\phi}.
\end{eqnarray}%
so that
\begin{eqnarray}
d\sigma ^{1} &=&\sigma ^{2}\wedge \sigma ^{3},  \notag \\
d\sigma ^{2} &=&\sigma ^{1}\wedge \sigma ^{3},  \notag \\
d\sigma ^{3} &=&-\sigma ^{1}\wedge \sigma ^{2},  \notag \\
\mathcal{L}_{L}\sigma ^{i} &=&0.
\end{eqnarray}%
Hence the 3-manifold with metric ${\frac{1}{4}}\left( (\sigma
^{1})^{2}+(\sigma ^{2})^{2}+(\sigma ^{3})^{2}\right) $ is either $\mathbb{S}%
^{3}$, the Nil-manifold or $\mathbb{H}^{3}$ according as to whether $c\theta
>0$, $\theta =0$ or $c\theta <0$ respectively.

\subsubsection{Solutions with $\func{Im}\protect\lambda =\func{Im}\protect%
\sigma =0$, $\protect\lambda \neq 0,\protect\sigma \neq 0$}

If $\lambda $ and $\sigma $ are real but non-vanishing, then $\psi =L$, and
from (\ref{lam}) and (\ref{sig}) we obtain

\begin{equation}
V_{I}X^{I}=\frac{1}{3\sqrt{2}\chi }\left( {\frac{f^{4}\theta }{K^{2}}}+{%
\frac{\sqrt{2}cf^{2}}{K^{2}}}\xi ^{2}-{\frac{c}{\sqrt{2}f^{2}}}\right) .
\end{equation}%
Moreover, from ({\ref{eqn:dfxcon1}}) we obtain
\begin{eqnarray}
d(\frac{X_{I}}{f^{2}}) &=&-\frac{\sqrt{2}}{f^{2}K^{2}}{c}X_{I}L,  \notag \\
d(\frac{1}{f^{2}}) &=&-\frac{dK^{2}}{f^{2}K^{2}}\ .
\end{eqnarray}%
This implies that $d(K^{2}f^{-2})=0$, and hence without loss of generality
we can set $K^{2}=f^{2}$ and the scalars are therefore constants.
Furthermore, we also find that

\begin{equation}
J=d\left( -{\frac{f^{2}\phi }{\sqrt{2}c\xi ^{2}}}\right) ,
\end{equation}
and

\begin{equation}
d\Omega =-d\left( \left( {\frac{1}{\sqrt{2}f^{2}}}+\frac{1}{2\sqrt{2}%
f^{4}\xi ^{2}}+ \frac{\theta }{2c\xi ^{2}} \right) \phi \right)
\end{equation}
with

\begin{equation}
F^{I}=d\left( f^{2}X^{I}(dt+\Omega )+{\frac{3\sqrt{2}\chi }{c\xi ^{2}}}
(X^{I}X^{J}-{\frac{1}{2}}Q^{IJ})V_{J}\phi +{\frac{1}{\sqrt{2}f^{2}\xi ^{2}}}
X^{I}\phi \right) ,
\end{equation}
and hence we can work in a gauge for which

\begin{equation}
\omega _{p,mn}\epsilon ^{mn}=-\frac{3\chi }{\sqrt{2}\xi ^{2}}\left( {\frac{%
6\chi }{c}}(X^{I}X^{J}-{\frac{1}{2}}Q^{IJ})V_{I}V_{J}+\frac{{1}}{f^{2}}{%
V_{I}X^{I}}\right) \phi _{p}.
\end{equation}%
It is then straightforward to prove the following identities:

\begin{eqnarray}
d\left( {\frac{\phi }{\xi ^{2}}}\right) &=&-{\frac{\sqrt{2}c}{f^{2}\xi ^{2}}}
\hat{L}\wedge \hat{\phi},  \notag \\
d\left( {\frac{\hat{L}}{f\xi }}\right) &=&- {\frac{1}{f \xi ^{3}}}({\theta }+%
{\frac{9\sqrt{2}\chi ^{2}}{c}}(X^{I}X^{J}- {\frac{1}{2}}Q^{IJ})V_{I}V_{J})%
\phi \wedge \hat{\phi},  \notag \\
d\left( {\frac{\hat{\phi}}{f\xi }}\right) &=&- {\frac{1}{f\xi ^{3}}}({\theta
}+{\frac{9\sqrt{2}\chi ^{2}}{c}}(X^{I}X^{J}- {\frac{1}{2}}Q^{IJ})V_{I}V_{J})%
\hat{L}\wedge \phi \ .  \notag \\
&&
\end{eqnarray}

There are then three cases to consider.

\begin{itemize}
\item[i)] If $c\theta +9\sqrt{2}\chi ^{2}(X^{I}X^{J}-{\frac{1}{2}}
Q^{IJ})V_{I}V_{J}<0$ then define

\begin{eqnarray}
\sigma ^{1} &=&-{\frac{\theta +{\frac{9\sqrt{2}\chi ^{2}}{c}}(X^{I}X^{J}-{%
\frac{1}{2}}Q^{IJ})V_{I}V_{J}}{\xi ^{2}}}\phi ,  \notag \\
\sigma ^{2} &=&{\frac{\sqrt{-\sqrt{2}(c\theta +9\sqrt{2}\chi ^{2}(X^{I}X^{J}-%
{\frac{1}{2}}Q^{IJ})V_{I}V_{J})}}{f\xi }}\hat{L},  \notag \\
\sigma ^{3} &=&{\frac{\sqrt{-\sqrt{2}(c\theta +9\sqrt{2}\chi ^{2}(X^{I}X^{J}-%
{\frac{1}{2}}Q^{IJ})V_{I}V_{J})}}{f\xi }}\hat{\phi}\ .
\end{eqnarray}
\end{itemize}

It is then straightforward to show that
\begin{eqnarray}
d\sigma ^{i} &=&-{\frac{1}{2}}\epsilon _{ijk}\sigma ^{j}\wedge \sigma ^{k},
\notag \\
\mathcal{L}_{L}\sigma ^{i} &=&0.
\end{eqnarray}

\begin{itemize}
\item[ii)] If $c\theta +9\sqrt{2}\chi ^{2}(X^{I}X^{J}-{\frac{1}{2}}
Q^{IJ})V_{I}V_{J}=0$ then define
\end{itemize}

\begin{eqnarray}
\sigma ^{1} &=&-{\frac{\phi }{\sqrt{2}c\xi ^{2}},}  \notag \\
\sigma ^{2} &=&{\frac{\hat{L}}{f\xi },}  \notag \\
\sigma ^{3} &=&{\frac{\hat{\phi}}{f\xi },}
\end{eqnarray}%
where
\begin{eqnarray}
d\sigma ^{1} &=&\sigma ^{2}\wedge \sigma ^{3},\qquad d\sigma ^{2}=d\sigma
^{3}=0,  \notag \\
\mathcal{L}_{L}\sigma ^{i} &=&0.
\end{eqnarray}

\begin{itemize}
\item[iii)] If $c\theta +9\sqrt{2}\chi ^{2}(X^{I}X^{J}-{\frac{1}{2}}
Q^{IJ})V_{I}V_{J}>0$ then define
\end{itemize}

\begin{eqnarray}
\sigma ^{1} &=&-{\frac{\theta +{\frac{9\sqrt{2}\chi ^{2}}{c}}(X^{I}X^{J}-{%
\frac{1}{2}}Q^{IJ})V_{I}V_{J}}{\xi ^{2}}}\phi ,  \notag \\
\sigma ^{2} &=&{\frac{\sqrt{\sqrt{2}(c\theta +9\sqrt{2}\chi ^{2}(X^{I}X^{J}-{%
\frac{1}{2}}Q^{IJ})V_{I}V_{J})}}{f\xi }}\hat{L},  \notag \\
\sigma ^{3} &=&{\frac{\sqrt{\sqrt{2}(c\theta +9\sqrt{2}\chi ^{2}(X^{I}X^{J}-{%
\frac{1}{2}}Q^{IJ})V_{I}V_{J})}}{f\xi }}\hat{\phi},
\end{eqnarray}

so that
\begin{eqnarray}
d\sigma ^{1} &=&\sigma ^{2}\wedge \sigma ^{3},  \notag \\
d\sigma ^{2} &=&\sigma ^{1}\wedge \sigma ^{3},  \notag \\
d\sigma ^{3} &=&-\sigma ^{1}\wedge \sigma ^{2},  \notag \\
\mathcal{L}_{L}\sigma ^{i} &=&0.
\end{eqnarray}
Hence the 3-manifold with metric ${\frac{1}{4}}\left( (\sigma
^{1})^{2}+(\sigma ^{2})^{2}+(\sigma ^{3})^{2}\right) $ is either $\mathbb{S}%
^{3}$, the Nil-manifold or $\mathbb{H}^{3}$ according as to whether $c\theta
+9\sqrt{2}\chi ^{2}(X^{I}X^{J}-{\frac{1}{2}}Q^{IJ})V_{I}V_{J}<0$, $c\theta
+9 \sqrt{2}\chi ^{2}(X^{I}X^{J}-{\frac{1}{2}}Q^{IJ})V_{I}V_{J}=0$ or $%
c\theta +9 \sqrt{2}\chi ^{2}(X^{I}X^{J}-{\frac{1}{2}}Q^{IJ})V_{I}V_{J}>0$
respectively.

\subsubsection{Solutions with $\protect\lambda =\protect\sigma =0$}

If $\lambda =\sigma =0$, then $f$ and the scalars $X^{I}$ are constant and $%
K^{2}$ is constant. Without loss of generality, we set $f=K^{2}=1$. The
following constraints also hold:
\begin{equation}
\theta =-{\frac{c}{\sqrt{2}}},\qquad 3\chi V_{I}X^{I}=-c
\end{equation}
so that $\theta \neq 0,V_{I}X^{I}\neq 0$ for these solutions. Furthermore,
we also find
\begin{equation}
d\Omega =-cK\wedge Z,
\end{equation}
and the gauge field strengths are given by
\begin{equation}
F^{I}=d\left( f^{2}X^{I}(dt+\Omega )\right) +6\chi V_{J}(X^{I}X^{J}- {\frac{1%
}{2}}Q^{IJ})(K\wedge Z-J)+cX^{I}(2K\wedge Z-J)  \label{fieldstv3}
\end{equation}
Note also that $K$ and $Z$ satisfy
\begin{eqnarray}
\nabla _{{\tilde{\mu}}}K_{{\tilde{\nu}}} &=&-c\Omega _{{\tilde{\mu}}}Z_{{%
\tilde{\nu}}},\quad  \notag \\
\nabla _{{\tilde{\mu}}}Z_{{\tilde{\nu}}} &=&c\Omega _{{\tilde{\mu}}}K_{{%
\tilde{\nu}}}
\end{eqnarray}
where here $\nabla $ denotes the covariant derivative restricted to the base
space, and ${\tilde{\mu}}$, ${\tilde{\nu}}$ are base space indices.

It is convenient to define

\begin{eqnarray}
{\hat{K}}_{p} &=&iK_{p},\quad {\hat{K}}_{\bar{p}}=-iK_{\bar{p}}  \notag \\
{\hat{Z}}_{p} &=&iZ_{p},\quad {\hat{Z}}_{\bar{p}}=-iZ_{\bar{p}}.
\end{eqnarray}
Then
\begin{equation}
J=K\wedge Z-{\hat{K}}\wedge {\hat{Z},}
\end{equation}
and
\begin{equation}
\nabla _{{\tilde{\mu}}}{\hat{K}}_{{\tilde{\nu}}}=-c{\hat{\Omega}}_{{\tilde{%
\mu}}}{\hat{Z}}_{{\tilde{\nu}}},\quad \nabla _{{\tilde{\mu}}} {\hat{Z}}_{{%
\tilde{\nu}}}=c{\hat{\Omega}}_{{\tilde{\mu}}}{\hat{K}}_{{\tilde{\nu}}}
\label{dhauxv1}
\end{equation}
where ${\hat{\Omega}}$ is defined by
\begin{equation}
{\hat{\Omega}}_{p}=\Omega _{p}+{\frac{1}{c}}\omega _{p,mn}\epsilon
^{mn},\quad {\hat{\Omega}}_{\bar{p}}=({\hat{\Omega}}_{p})^{\ast }.
\end{equation}
Note that ({\ref{dhauxv1}}) implies that

\begin{equation}
d{\hat{\Omega}}\wedge {\hat{K}}=d{\hat{\Omega}}\wedge {\hat{Z}}=0,
\end{equation}
and hence
\begin{equation}
d{\hat{\Omega}}=\Psi {\hat{K}}\wedge {\hat{Z}}
\end{equation}
where $\Psi $ is fixed by comparing the integrability condition associated
with ({\ref{eqn:killpot1}}) with the expression for the gauge field
strengths ({\ref{fieldstv3}}). We find
\begin{equation}
\Psi ={\frac{1}{c}}(9\chi ^{2}Q^{IJ}V_{I}V_{J}-c^{2}).
\end{equation}
Next we define

\begin{eqnarray}
A_{p} &=&\cos ctK_{p}+\sin ctZ_{p},\quad  \notag \\
A_{\bar{p}} &=&\cos ctK_{\bar{p}}+\sin ctZ_{\bar{p}},  \notag \\
B_{p} &=&\epsilon _{pq}A^{q}=\cos ctZ_{p}-\sin ctK_{p},  \notag \\
B_{\bar{p}} &=&\cos ctZ_{\bar{p}}-\sin ctK_{\bar{p}},
\end{eqnarray}
so that
\begin{equation}
\partial_{t}A=\partial_{t}B=0.
\end{equation}
We also define
\begin{eqnarray}
{\hat{A}}_{p} &=&iA_{p},\quad {\hat{A}}_{\bar{p}}=-iA_{\bar{p}},  \notag \\
{\hat{B}}_{p} &=&iB_{p},\quad {\hat{B}}_{\bar{p}}=-iB_{\bar{p}} \ .
\end{eqnarray}
Then $A,B,{\hat{A}},{\hat{B}}$ form an orthonormal time-independent basis
for the K\"{a}hler base space, such that

\begin{eqnarray}
\nabla _{{\tilde{\mu}}}A_{{\tilde{\nu}}} &=&-c\Omega _{{\tilde{\mu}}}B_{{%
\tilde{\nu}}},\quad  \notag \\
\nabla _{{\tilde{\mu}}}B_{{\tilde{\nu}}} &=&c\Omega _{{\tilde{\mu}}}A_{{%
\tilde{\nu}}},\quad  \notag \\
d\Omega &=&-cA\wedge B
\end{eqnarray}
and

\begin{eqnarray}
\nabla _{{\tilde{\mu}}}{\hat{A}}_{{\tilde{\nu}}} &=&-c{\hat{\Omega}}_{{%
\tilde{\mu}}}{\hat{B}}_{{\tilde{\nu}}},\quad  \notag \\
\nabla _{{\tilde{\mu}}}{\hat{B}}_{{\tilde{\nu}}} &=&c{\hat{\Omega}}_{{\tilde{%
\mu}}}{\hat{A}}_{{\tilde{\nu}}},\quad  \notag \\
d{\hat{\Omega}} &=&{\frac{1}{c}}(9\chi ^{2}Q^{IJ}V_{I}V_{J}-c^{2}){\hat{A}}
\wedge {\hat{B}.}
\end{eqnarray}
Now note that

\begin{equation}
\lbrack A,B]=(ci_{A}\Omega )A+(ci_{B}\Omega )B.
\end{equation}
It therefore follows that there exist functions $\alpha _{1},\alpha
_{2},\beta _{1},\beta _{2}$ and co-ordinates $w_{1},w_{2}$ such that
\begin{equation}
A=\alpha _{1}{\frac{\partial }{\partial w_{1}}}+\alpha _{2}{\frac{\partial
} {\partial w_{2}}},\quad B=\beta _{1}{\frac{\partial }{\partial w_{1}}}%
+\beta _{2}{\frac{\partial }{\partial w_{2}}.}
\end{equation}
Suppose that the remaining co-ordinates on the base are $y^{1},y^{2}$. As ${%
\hat{A}}$ and ${\hat{B}}$ are orthogonal to $A,B$, there exist functions $%
\rho _{i},\nu _{i}$ such that
\begin{equation}
{\hat{A}}=\rho _{i}dy^{i},\quad {\hat{B}}=\nu _{i}dy^{i}
\end{equation}
for $i=1,2$.

Similarly, as
\begin{equation}
\lbrack {\hat{A}},{\hat{B}}]=(ci_{{\hat{A}}}{\hat{\Omega}}){\hat{A}}+ (ci_{{%
\hat{B}}}{\hat{\Omega}}){\hat{B}}
\end{equation}
it also follows that there exist functions ${\hat{\rho}}_{i}$, ${\hat{\nu}}%
_{i}$ and $x^{i}$ for $i=1,2$ such that
\begin{equation}
A={\hat{\rho}}_{i}dx^{i},\quad B={\hat{\nu}}_{i}dx^{i}
\end{equation}
for $i=1,2$. As $A,B,{\hat{A}},{\hat{B}}$ form an orthonormal basis for the
base space, we can without loss of generality take $x^{1},x^{2},y^{1},y^{2}$
to be co-ordinates on the base space.

In principle, the functions $\rho_i, {\hat{\rho}}_i, \nu_i, {\hat{\nu}}_i$
can depend on all the co-ordinates. However, note that

\begin{equation}
J=A\wedge B-{\hat{A}}\wedge {\hat{B}}=({\hat{\rho}}_{1}{\hat{\nu}}_{2}- {%
\hat{\nu}}_{1}{\hat{\rho}}_{2})dx^{1}\wedge dx^{2}-(\rho _{1}\nu _{2}-\nu
_{1}\rho _{2})dy^{1}\wedge dy^{2}.
\end{equation}
Imposing the constraint $dJ=0$ thus gives the conditions

\begin{equation}
{\frac{\partial }{\partial y^{i}}}({\hat{\rho}}_{1}{\hat{\nu}}_{2}- {\hat{\nu%
}}_{1}{\hat{\rho}}_{2})={\frac{\partial }{\partial x^{i}}}(\rho _{1}\nu
_{2}-\nu _{1}\rho _{2})=0.
\end{equation}
One therefore can set

\begin{equation}
\Omega =d\Phi +\Omega _{T}
\end{equation}
with $\Omega _{T}=\Omega _{Ti}(x^{1},x^{2})dx^{i}$ satisfying
\begin{equation}
d\Omega _{T}=-c({\hat{\rho}}_{1}{\hat{\nu}}_{2} -{\hat{\nu}}_{1}{\hat{\rho}}%
_{2})dx^{1}\wedge dx^{2}.
\end{equation}
We also set
\begin{equation}
{\hat{\Omega}}=d{\hat{\Phi}}+{\hat{\Omega}}_{T}
\end{equation}
where ${\hat{\Omega}}_{T}={\hat{\Omega}}_{Ti}(y^{1},y^{2})dy^{i}$ satisfies
\begin{equation}
d{\hat{\Omega}}_{T}={\frac{1}{c}}(9\chi ^{2}Q^{IJ}V_{I}V_{J}-c^{2})(\rho
_{1}\nu _{2}-\nu _{1}\rho _{2})dy^{1}\wedge dy^{2}.
\end{equation}
Here $\Phi $ and ${\hat{\Phi}}$ are functions of $x^{i},y^{i}$. Next we
define
\begin{eqnarray}
A^{\prime } &=&\cos c\Phi A+\sin c\Phi B,  \notag \\
B^{\prime } &=&-\sin c\Phi A+\cos c\Phi B,  \notag \\
{\hat{A}}^{\prime } &=&\cos c{\hat{\Phi}\hat{A}}+\sin c{\hat{\Phi}}{\hat{B},}
\notag \\
{\hat{B}}^{\prime } &=&-\sin c{\hat{\Phi}}{\hat{A}}+\cos c{\hat{\Phi}} {\hat{%
B}.}
\end{eqnarray}
Note that $A^{\prime },B^{\prime },{\hat{A}}^{\prime },{\hat{B}}^{\prime }$
are an orthonormal basis of the K\"{a}hler base with the property that

\begin{equation}
\nabla _{{\tilde{\mu}}}A^{\prime }{}_{{\tilde{\nu}}}=-c\Omega _{T{\tilde{\mu}%
}}B^{\prime }{}_{{\tilde{\nu}}},\quad \nabla _{{\tilde{\mu}}}B^{\prime }{}_{{%
\tilde{\nu}}}=c\Omega _{T{\tilde{\mu}}}A^{\prime }{}_{{\tilde{\nu}}}
\end{equation}
and

\begin{equation}
\nabla _{{\tilde{\mu}}}{{\hat{A}}^{\prime }}_{{\tilde{\nu}}}=-c{\hat{\Omega}}%
_{T{\tilde{\mu}}}{{\hat{B}}^{\prime }}_{{\tilde{\nu}}},\quad \nabla _{{%
\tilde{\mu}}}{{\hat{B}}^{\prime }}_{{\tilde{\nu}}}=c{\hat{\Omega}}_{T{\tilde{%
\mu}}} {{\hat{A}}^{\prime }}_{{\tilde{\nu}}}
\end{equation}
with

\begin{equation}
d\Omega _{T}=-cA^{\prime }\wedge B^{\prime },\quad d{\hat{\Omega}}_{T}= {%
\frac{1}{c}}(9\chi ^{2}Q^{IJ}V_{I}V_{J}-c^{2}){\hat{A}}^{\prime }\wedge {%
\hat{B}}^{\prime }.
\end{equation}
These constraints therefore imply that

\begin{equation}
\mathcal{L}_{\frac{\partial }{\partial y^{i}}}A^{\prime }=\mathcal{L}_{\frac{%
\partial }{\partial y^{i}}}B^{\prime }=\mathcal{L}_{\frac{\partial }{%
\partial x^{i}}}{\hat{A}}^{\prime }=\mathcal{L}_{\frac{\partial }{\partial
x^{i}}}{\hat{B}}^{\prime }=0.
\end{equation}%
Hence the K\"{a}hler base is a product of two 2-manifolds $M_{1}$, $M_{2}$,
with metric
\begin{equation}
ds_{\mathcal{B}}^{2}=ds^{2}(M_{1})+ds^{2}(M_{2}).
\end{equation}%
Taking the orthonormal basis $\mathbf{e}^{1}=A^{\prime },\mathbf{e}%
^{2}=B^{\prime },\mathbf{e}^{3}={\hat{A}}^{\prime },\mathbf{e}^{4}={\hat{B}}%
^{\prime }$, the metrics on $M_{1}$ and $M_{2}$ are
\begin{equation}
ds^{2}(M_{1})=(\mathbf{e}^{1})^{2}+(\mathbf{e}^{2})^{2},\quad ds^{2}(M_{2})=(%
\mathbf{e}^{3})^{2}+(\mathbf{e}^{4})^{2}\ .
\end{equation}%
It is then straightforward to compute the curvature in this basis. We find
that the only non-vanishing components are fixed by

\begin{equation}
R_{1212}=-c^{2},\quad R_{3434}=9\chi ^{2}Q^{IJ}V_{I}V_{J}-c^{2}.
\end{equation}
Therefore we conclude that $M_{1}$ is $\mathbb{H}^{2}$, and $M_{2}$ is $%
\mathbb{H}^{2}$, $\mathbb{R}^{2}$ or $\mathbb{S}^{2}$ depending on whether $%
9\chi ^{2}Q^{IJ}V_{I}V_{J}-c^{2}<0$, $9\chi ^{2}Q^{IJ}V_{I}V_{J}-c^{2}=0$ or
$9\chi ^{2}Q^{IJ}V_{I}V_{J}-c^{2}>0$ respectively.

\subsection{Solutions with $c_{2}=0$ and $c_{3}{}^{2}+c_{4}{}^{2}=\protect%
\varrho ^{2}\neq 0$}

In the next case, we shall assume that $c_{3}$ and $c_{3}$ do not both
vanish, and define the vector fields $W$, $Y$ on the K\"{a}hler base via

\begin{eqnarray}
W &=&c_{4}K-c_{3}Z,  \notag \\
Y &=&c_{3}K+c_{4}Z.
\end{eqnarray}
Then note that $W\neq 0$ and $Y\neq 0$, and ({\ref{conf1}}), ({\ref{conf2}})
and ({\ref{conf3}}) can be rewritten in terms of $Y$ and $W$ as

\begin{eqnarray}
\nabla _{p}W_{\bar{q}} &=&0,  \notag \\
\nabla _{p}W_{q} &=&{\frac{\varrho ^{2}}{\sqrt{2}}}\epsilon _{pq},  \notag \\
\nabla _{p}Y_{\bar{q}} &=&{\frac{\varrho ^{2}}{\sqrt{2}}}\delta _{p\bar{q}},
\notag \\
\nabla _{p}Y_{q} &=&0.
\end{eqnarray}
Therefore we find that $W$ is a holomorphic Killing vector on the base and
satisfies

\begin{equation}
dW=-\sqrt{2}\varrho ^{2}J.
\end{equation}
Moreover, $W$ preserves the complex structure

\begin{equation}
\mathcal{L}_{W}J=0.
\end{equation}
In contrast, $Y$ defines a closed 1-form on the base, which is conformally
Killing with

\begin{equation}
\mathcal{L}_{Y}h=\sqrt{2}\varrho ^{2}h,\qquad \mathcal{L}_{Y}J=\sqrt{2}
\varrho ^{2}J.
\end{equation}
Here $h$ denotes the metric of the K\"{a}hler base. From ({\ref{eqn:commutx1}%
}) it is clear that $W$ and $Y$ commute, so that locally one can choose
co-ordinates $\phi $, $\psi $ so that
\begin{equation}
W={\frac{\partial }{\partial \phi }},\quad Y={\frac{\partial }{\partial \psi
}}
\end{equation}
and let the remaining two co-ordinates of the base space be $x^{1}$, $x^{2}$.

Note that $Y$ and $W$ satisfy

\begin{equation}
h(Y,W)=0,\qquad Y^{2}=W^{2}.
\end{equation}
Then, one can write the metric on the K\"{a}hler base locally as

\begin{equation}
ds^{2}=e^{\sqrt{2}\varrho ^{2}\psi }\big[S^{2}(d\phi +\chi
_{1})^{2}+S^{2}(d\psi +\chi _{2})^{2}+T^{2}((dx^{1})^{2}+(dx^{2})^{2})\big]
\label{kahlmet}
\end{equation}
where $S=S(x^{1},x^{2})$, $T=T(x^{1},x^{2})$ and $\chi _{i}=\chi
_{ij}(x^{1},x^{2})dx^{j}$. By making a co-ordinate transformation of the
form
\begin{eqnarray}
\psi &=&\psi ^{\prime }-{\frac{1}{\sqrt{2}\varrho ^{2}}}\log S^{2},  \notag
\\
\phi &=&\phi ^{\prime },  \notag \\
x^{1} &=&(x^{1})^{\prime },  \notag \\
x^{2} &=&(x^{2})^{\prime },
\end{eqnarray}
one can without loss of generality set $S=1$ in ({\ref{kahlmet}}) and drop
primes throughout.

Then it is straightforward to show that the condition $dY=0$ implies that $%
\chi_2=0$, so that

\begin{equation}
ds^{2}=e^{\sqrt{2}\varrho ^{2}\psi }\left[ (d\phi +\beta )^{2}+d\psi
^{2}+T^{2}((dx^{1})^{2}+(dx^{2})^{2})\right]  \label{simplerbasemet1}
\end{equation}
where $\beta =\beta _{i}(x^{1},x^{2})dx^{i}$, and

\begin{equation}
J=-{\frac{1}{\sqrt{2}\varrho ^{2}}}d\left( e^{\sqrt{2}\varrho ^{2}\psi
}(d\phi +\beta )\right) .
\end{equation}
The necessary and sufficient condition in order for $J$ to be a covariantly
constant complex structure is

\begin{equation}
T^{2}={\frac{1}{\sqrt{2}\varrho ^{2}}}|({\frac{\partial \beta _{2}}{\partial
x^{1}}}-{\frac{\partial \beta _{1}}{\partial x^{2}}})|
\end{equation}
In fact, we can take ${\frac{\partial \beta _{2}}{\partial x^{1}}}- {\frac{%
\partial \beta _{1}}{\partial x^{2}}}>0$ without loss of generality, (this
can be obtained, if necessary, by making the re-definition $\beta
_{2}\rightarrow -\beta _{2}$ and $x^{2}\rightarrow -x^{2}$). So

\begin{equation}
T^{2}={\frac{1}{\sqrt{2}\varrho ^{2}}}\left( {\frac{\partial \beta _{2}} {%
\partial x^{1}}}-{\frac{\partial \beta _{1}}{\partial x^{2}}}\right) .
\end{equation}
In addition, ({\ref{nextsimp6}}) and ({\ref{nextsimp7}}) can be rewritten as

\begin{eqnarray}
\frac{1}{f}c_{3}-\frac{2\sqrt{2}}{f^{2}}K^{p}\nabla _{p}f+2i\chi V_{I}X^{I}(
\func{Im}\sigma ) &=&0,  \notag \\
\frac{1}{f}c_{4}-\frac{2\sqrt{2}}{f^{2}}Z^{\bar{p}}\nabla _{\bar{p}}f-2i\chi
V_{I}X^{I}(\func{Im}\lambda ) &=&0.  \label{tsim1}
\end{eqnarray}
The real portions of ({\ref{tsim1}}) imply that

\begin{equation}
\mathcal{L}_{K}f={\frac{c_{3}}{\sqrt{2}}}f,\text{ \ \ }\mathcal{L}_{Z}f= {%
\frac{c_{4}}{\sqrt{2}}}f
\end{equation}
and hence

\begin{equation}
\mathcal{L}_{W}f=0,\text{ \ \ \ }\mathcal{L}_{Y}f={\frac{1}{\sqrt{2}}}
\varrho ^{2}f.
\end{equation}
Therefore

\begin{equation}
f=e^{{\frac{1}{\sqrt{2}}}\varrho ^{2}\psi }u(x^{1},x^{2})
\end{equation}
for some function $u(x^{1},x^{2})$. It should be noted that although the K%
\"{a}hler metric $h$ has a conformal dependence on $\psi $, the portion of
the metric $f^{-2}h$ which appears in the five dimensional metric does not
depend on either $\phi $ or $\psi $.

In order to examine the behaviour of the scalars, note that ({\ref{scon}})
implies that

\begin{equation}
X^{I}=X^{I}(x^{1},x^{2}).
\end{equation}
To proceed further, we introduce the following holomorphic basis for the K%
\"{a}hler base space

\begin{eqnarray}
\mathbf{e}^{1} &=&{\frac{e^{{\frac{1}{\sqrt{2}}}\varrho ^{2}\psi }}{\sqrt{2}%
\varrho }}\left( c_{3}d\psi +c_{4}(d\phi +\beta )-i\varrho Tdx^{1}\right) ,
\notag \\
\mathbf{e}^{2} &=&{\frac{e^{{\frac{1}{\sqrt{2}}}\varrho ^{2}\psi }}{\sqrt{2}%
\varrho }}\left( -c_{4}d\psi +c_{3}(d\phi +\beta )+i\varrho Tdx^{2}\right)
\end{eqnarray}%
with $\mathbf{e}^{\bar{1}}$, $\mathbf{e}^{\bar{2}}$ obtained by complex
conjugation. It is straightforward to show that in this basis
\begin{equation}
J=-\left( \mathbf{e}^{1}\wedge \mathbf{e}^{2}+\mathbf{e}^{\bar{1}}\wedge
\mathbf{e}^{\bar{2}}\right)
\end{equation}%
as expected, and
\begin{eqnarray}
K^{1} &=&K^{\bar{1}}={\frac{1}{\sqrt{2}\varrho }}e^{{\frac{1}{\sqrt{2}}}%
\varrho ^{2}\psi },  \notag \\
K^{2} &=&K^{\bar{2}}=0,  \notag \\
Z^{1} &=&Z^{\bar{1}}=0,  \notag \\
Z^{2} &=&Z^{\bar{2}}=-{\frac{1}{\sqrt{2}\varrho }}e^{{\frac{1}{\sqrt{2}}}%
\varrho ^{2}\psi }.
\end{eqnarray}%
Using this basis, one can compute explicitly the following components of the
spin connection:

\begin{eqnarray}
\omega _{1,mn}\epsilon ^{mn} &=&e^{-{\frac{1}{\sqrt{2}}}\varrho ^{2}\psi
}\left( -\varrho c_{4}+{\frac{i}{\sqrt{2}T^{2}}\frac{\partial {T}}{\partial
x^{2}}}\right) ,  \notag \\
\omega _{2,mn}\epsilon ^{mn} &=&e^{-{\frac{1}{\sqrt{2}}}\varrho ^{2}\psi
}\left( -\varrho c_{3}+{\frac{i}{\sqrt{2}T^{2}}\frac{\partial {T}}{\partial
x^{1}}}\right) .
\end{eqnarray}

\bigskip

\bigskip

Moreover, it is straightforward to show that the imaginary portion of ({\ref%
{tsim1}}) implies that

\begin{equation}
\sqrt{2}\chi V_{I}X^{I}\func{Im}\sigma ={\frac{1}{\varrho Tu^{2}}}e^{-{\frac{%
1}{\sqrt{2}}}\varrho ^{2}\psi }\frac{\partial u}{\partial x^{1}}
\label{imsig1}
\end{equation}%
and
\begin{equation}
\sqrt{2}\chi V_{I}X^{I}\func{Im}\lambda ={\frac{1}{\varrho Tu^{2}}}e^{-{%
\frac{1}{\sqrt{2}}}\varrho ^{2}\psi }{\frac{\partial u}{\partial x^{2}}}\ .
\label{imsig2}
\end{equation}%
The imaginary parts of ({\ref{nextsimp4}}) and ({\ref{nextsimp5}}) can then
be rewritten as

\begin{eqnarray}
{\frac{\partial }{\partial x^{1}}}\left( \frac{X_{I}}{u^{2}}\right) &=&-2%
\sqrt{2}\chi \frac{\varrho }{u}e^{{\frac{1}{\sqrt{2}}}\varrho ^{2}\psi
}TV_{I}\func{Im}\sigma ,\text{ }  \notag \\
\text{\ }\frac{\partial }{\partial x^{2}}\left( \frac{X_{I}}{u^{2}}\right)
&=&-2\sqrt{2}\chi \frac{\varrho }{u}e^{{\frac{1}{\sqrt{2}}}\varrho ^{2}\psi
}TV_{I}\func{Im}\lambda .  \label{xsimpe1}
\end{eqnarray}%
Note that as the $V_{I}$ do not all vanish in the gauged theory, it follows
that the imaginary parts of $\lambda $ and $\sigma $ do not depend on $\phi $%
, and depend on $\psi $ via the factor $e^{-{\frac{1}{\sqrt{2}}}\varrho
^{2}\psi }$. It is therefore convenient to define

\begin{eqnarray}
\mathcal{G}(x^{1},x^{2}) &=&\frac{2i}{u}e^{{\frac{1}{\sqrt{2}}}\varrho
^{2}\psi }\func{Im}\lambda  \notag \\
\mathcal{H}(x^{1},x^{2}) &=&\frac{2i}{u}e^{{\frac{1}{\sqrt{2}}}\varrho
^{2}\psi }\func{Im}\sigma ,
\end{eqnarray}%
and rewrite ({\ref{imsig1}}), ({\ref{imsig2}}) and ({\ref{xsimpe1}}) as

\begin{equation}
\chi V_{I}X^{I}\mathcal{H}={\frac{\sqrt{2}i}{\varrho Tu^{3}}\frac{\partial u%
}{\partial x^{1}},}  \label{imsig1b}
\end{equation}

\begin{equation}
\chi V_{I}X^{I}\mathcal{G}={\frac{\sqrt{2}i}{\varrho Tu^{3}}\frac{\partial u%
}{\partial x^{2}},}  \label{imsig2b}
\end{equation}%
and

\begin{eqnarray}
{\frac{\partial }{\partial x^{1}}}(\frac{X_{I}}{u^{2}}) &=&\sqrt{2}i\chi
\varrho T\mathcal{H}V_{I},  \notag \\
\frac{\partial }{\partial x^{2}}(\frac{X_{I}}{u^{2}}) &=&\sqrt{2}i\chi
\varrho T\mathcal{G}V_{I}\ .  \label{xsimpe1b}
\end{eqnarray}%
We next consider the constraints ({\ref{nextsimp9}}), ({\ref{nextsim12}})
and ({\ref{nextsim13}}). These are equivalent to

\begin{equation}
{\frac{\partial }{\partial x^{2}}}(T\mathcal{H})={\frac{\partial }{\partial
x^{1}}}(T\mathcal{G}),  \label{difconstr1}
\end{equation}

and

\begin{equation}
{\frac{\partial }{\partial x^{1}}}(\frac{\mathcal{H}}{T})={\frac{\partial } {%
\partial x^{2}}}(\frac{\mathcal{G}}{T}),\qquad {\frac{\partial }{\partial
x^{1}}}(\frac{\mathcal{G}}{T})=-{\frac{\partial }{\partial x^{2}}} (\frac{%
\mathcal{H}}{T}).  \label{difconstr2}
\end{equation}

Note that ({\ref{difconstr1}}) implies the integrability condition
associated with ({\ref{xsimpe1b}}), and ({\ref{difconstr2}}) implies that $%
T^{-1}\mathcal{H}$ and $T^{-1}\mathcal{G}$ satisfy the Cauchy-Riemann
equations. Hence, $T^{-1}(\mathcal{H}+i\mathcal{G})$ is a holomorphic
function of $x^{1}+ix^{2}$.

The components of $d\Omega $ are also fixed by ({\ref{nextsimp9}}), ({\ref%
{nextsim12}}) and ({\ref{nextsim13}}) to be

\begin{eqnarray}
d\Omega _{1\bar{2}} &=&3u^{-4}e^{-2\sqrt{2}\varrho ^{2}\psi }\chi V_{I}X^{I},
\notag \\
d\Omega _{1\bar{1}} &=&d\Omega _{2\bar{2}}=\varrho ^{2}e^{-2\sqrt{2}\varrho
^{2}\psi }(c_{4}\mathcal{H}-c_{3}\mathcal{G}),  \notag \\
d\Omega _{12} &=&e^{-2\sqrt{2}\varrho ^{2}\psi }\left[ \varrho ^{2}(c_{4}%
\mathcal{G}+c_{3}\mathcal{H})-{\frac{i}{\sqrt{2}}}\varrho \left( \frac{1}{T}{%
\frac{\partial \mathcal{G}}{\partial x^{2}}}+\frac{1}{T^{2}}\mathcal{H}{%
\frac{\partial T}{\partial x^{1}}}\right) \right]
\end{eqnarray}%
with the remaining components determined by complex conjugation; this
exhausts the content of ({\ref{nextsimp9}}), ({\ref{nextsim12}}) and ({\ref%
{nextsim13}}). Using these components, it is straightforward to compute $%
d\Omega $ in the co-ordinate basis; we find
\begin{eqnarray}
d\Omega &=&e^{-\sqrt{2}\varrho ^{2}\psi }\big[\left( -{\frac{i}{\sqrt{2}}}%
\varrho (\frac{1}{T}{\frac{\partial \mathcal{G}}{\partial x^{2}}}+\frac{%
\mathcal{H}}{T^{2}}{\frac{\partial T}{\partial x^{1}}})+3\frac{\chi }{u^{4}}%
V_{I}X^{I}\right) d\psi \wedge (d\phi +\beta )  \notag \\
&&+\left( -{\frac{i}{\sqrt{2}}}\varrho (T{\frac{\partial \mathcal{G}}{%
\partial x^{2}}}+\mathcal{H}{\frac{\partial T}{\partial x^{1}}})-3\chi \frac{%
T^{2}}{u^{4}}V_{I}X^{I}\right) dx^{1}\wedge dx^{2}  \notag \\
&&+iT\varrho ^{3}\left( d\psi \wedge (-\mathcal{G}dx^{1}+\mathcal{H}%
dx^{2})+(d\phi +\beta )\wedge (\mathcal{H}dx^{1}+\mathcal{G}dx^{2})\right) %
\big].  \label{domegafo1}
\end{eqnarray}

Using ({\ref{nextsimp8}}) and ({\ref{nextsimp8a}}), the gauge field
strengths for these solutions can be written as

\begin{equation}
F^{I}=d\left( f^{2}X^{I}(dt+\Omega )\right) +6\chi (X^{I}X^{J}-{\frac{1}{2}}%
Q^{IJ})V_{J}\frac{T^{2}}{u^{2}}dx^{1}\wedge dx^{2}
\end{equation}%
which satisfy $dF^{I}=0$ automatically.

Next consider the integrability condition associated with ({\ref%
{eqn:killpot1}}): this can be written as

\begin{equation}
3\chi V_{I}\big(F^{I}-d(f^{2}X^{I}(dt+\Omega ))\big)=-d(\omega
_{p,mn}\epsilon ^{mn}\mathbf{e}^{p}+\omega _{\bar{p},\bar{m}\bar{n}}\epsilon
^{\bar{m}\bar{n}}\mathbf{e}^{\bar{p}}).
\end{equation}%
This can be evaluated to give the constraint

\begin{equation}
\Box \log T+2\varrho ^{4}T^{2}=18\chi ^{2}(X^{I}X^{J}-{\frac{1}{2}}%
Q^{IJ})V_{I}V_{J}\frac{T^{2}}{u^{2}}
\end{equation}%
where $\Box =({\frac{\partial }{\partial x^{1}}})^{2}+({\frac{\partial }{%
\partial x^{2}}})^{2}$ is the Laplacian on $\mathbb{R}^{2}$. In fact, this
constraint enables ({\ref{domegafo1}}) to be solved for $\Omega $ (up to a
total derivative); we find

\begin{equation}
\Omega =-{\frac{e^{-\sqrt{2}\varrho ^{2}\psi }}{\sqrt{2}}}\left[ iT\varrho (-%
\mathcal{G}dx^{1}+\mathcal{H}dx^{2})+\left( -{\frac{i}{\sqrt{2}}} \frac{1}{%
\varrho }(\frac{1}{T}{\frac{\partial \mathcal{G}}{\partial x^{2}}} +\frac{%
\mathcal{H}}{T^{2}}{\frac{\partial T}{\partial x^{1}}})+\frac{3\chi }{%
\varrho^2 u^{2}} V_{I}X^{I}\right) (d\phi +\beta )\right]
\end{equation}
and the constraint ({\ref{xsimpe1b}}) implies that the scalars $X_{I}$ are
given by

\begin{equation}
X_{I}=u^{2}q_{I}+{\chi u}^{2}\left( -{\frac{i}{\sqrt{2}\varrho ^{3}}}(\frac{1%
}{T}{\frac{\partial \mathcal{G}}{\partial x^{2}}}+\frac{\mathcal{H}}{T^{2}}{%
\frac{\partial T}{\partial x^{1}}})+\frac{3\chi }{\varrho ^{4}u^{2}}%
V_{J}X^{J}\right) V_{I}
\end{equation}%
for constant $q_{I}$.

Lastly, consider the equations ({\ref{nextsimp10}}) and ({\ref{nextsimp11}}%
). These are equivalent to
\begin{equation}
\sqrt{2}{\tilde{d}}\left( c_{4}(\frac{1}{f}\func{Re}\lambda )-c_{3}(\frac{1}{%
f}\func{Re}\sigma )\right) =i_{W}d\Omega
\end{equation}%
and
\begin{equation}
\sqrt{2}{\tilde{d}}\left( c_{3}(\frac{1}{f}\func{Re}\lambda )+c_{4}(\frac{1}{%
f}\func{Re}\sigma )\right) =i_{Y}d\Omega +\sqrt{2}\varrho ^{2}\Omega \ .
\end{equation}%
where ${\tilde{d}}$ denotes the restriction of the exterior derivative to
hypersurfaces of constant $t.$ The integrability conditions of these two
equations are
\begin{equation}
\mathcal{L}_{W}d\Omega =0
\end{equation}%
and
\begin{equation}
\mathcal{L}_{Y}d\Omega =-\sqrt{2}\varrho ^{2}d\Omega
\end{equation}%
which hold automatically.

\subsection{Solutions with $c_2=c_3=c_4=0$}

We now turn to the class of solutions with $c_{2}=c_{3}=c_{4}=0.$ It is
clear that ({\ref{conf1}}), ({\ref{conf2}}) and ({\ref{conf3}}) imply that $%
K $ and $Z$ are covariantly constant. In particular, this implies that $%
K^{2} $ is constant and without loss of generality we can set $K^{2}=1$. One
can choose co-ordinates $\phi $, $\psi $ so that locally

\begin{equation}
K={\frac{\partial }{\partial \phi }},\quad Z={\frac{\partial }{\partial \psi}%
,}
\end{equation}
and two additional co-ordinates $x^{1}$, $x^{2}$ can be chosen on the base
so that the K\"{a}hler metric takes the form

\begin{equation}
ds^{2}=d\phi ^{2}+d\psi ^{2}+T^{2}((dx^{1})^{2}+(dx^{2})^{2})
\end{equation}
where $T=T(x^{1},x^{2})$.

Recall that $K\wedge Z-{\frac{1}{2}}J$ is anti-self-dual, so taking positive
orientation on the base with respect to $T^{2}d\phi \wedge d\psi \wedge
dx^{1}\wedge dx^{2}$, we have

\begin{equation}
J=d\phi \wedge d\psi -T^{2}dx^{1}\wedge dx^{2}.
\end{equation}
This complex structure is automatically covariantly constant. Then the real
portions of ({\ref{nextsimp6}}) and ({\ref{nextsimp7}}) imply that

\begin{equation}
\mathcal{L}_{K}f=\mathcal{L}_{Z}f=0,
\end{equation}
so that $f$ is only a function of $x^{1}$ and $x^{2}$. Also, as in the
previous case, the real part of ({\ref{nextsimp4}}) and ({\ref{nextsimp5}})
implies that

\begin{equation}
\mathcal{L}_{K}X^{I}=\mathcal{L}_{Z}X^{I}=0,
\end{equation}
so
\begin{equation}
X^{I}=X^{I}(x^{1},x^{2}).
\end{equation}
It is convenient to define
\begin{equation}
\mathcal{G}=\frac{2i}{f}\func{Im}\lambda ,\quad \mathcal{H}=\frac{2i}{f}
\func{Im}\sigma .
\end{equation}
Then the remaining portions of ({\ref{nextsimp6}}), ({\ref{nextsimp7}}), ({%
\ref{nextsimp4}}) and ({\ref{nextsimp5}}) can be rewritten as

\begin{eqnarray}
{\frac{\partial }{\partial x^{1}}}(\frac{X_{I}}{f^{2}}) &=&\sqrt{2}i\chi T
\mathcal{H}V_{I},  \notag \\
\frac{\partial }{\partial x^{2}}(\frac{X_{I}}{f^{2}}) &=&\sqrt{2}i\chi T
\mathcal{G}V_{I}.
\end{eqnarray}
As the $V_{I}$ do not all vanish, these constraints imply that
\begin{equation}
\mathcal{G}=\mathcal{G}(x^{1},x^{2}),\quad \mathcal{H}=\mathcal{H}
(x^{1},x^{2}).
\end{equation}
We take the following holomorphic basis for the K\"{a}hler base space:

\begin{eqnarray}
\mathbf{e}^{1} &=&{\frac{1}{\sqrt{2}}}(Tdx^{1}+id\phi ),  \notag \\
\mathbf{e}^{2} &=&{\frac{1}{\sqrt{2}}}(Tdx^{2}+id\psi )\ ,
\end{eqnarray}%
with $\mathbf{e}^{\bar{1}}$, $\mathbf{e}^{\bar{2}}$ fixed by complex
conjugation.

In this basis, we obtain the following spin connection components:

\begin{eqnarray}
\omega _{1,mn}\epsilon ^{mn} &=&{\frac{1}{\sqrt{2}T^{2}}}{\frac{\partial T} {%
\partial x^{2}},}  \notag \\
\omega _{2,mn}\epsilon ^{mn} &=&-{\frac{1}{\sqrt{2}T^{2}}}{\frac{\partial T%
} {\partial x^{1}}}
\end{eqnarray}
and the components of $K$ and $Z$ are:
\begin{eqnarray}
K_{1} &=&-K_{\bar{1}}=-{\frac{i}{\sqrt{2}}},\qquad K_{2}=K_{\bar{2}}=0,
\notag \\
Z_{1} &=&Z_{\bar{1}}=0,\qquad Z_{2}=-Z_{\bar{2}}=-{\frac{i}{\sqrt{2}}.}
\end{eqnarray}

To proceed, we turn to the constraint ({\ref{nextsimp9}}). This implies that
\begin{equation}
d\Omega _{1\bar{1}}=d\Omega _{2\bar{2}}
\end{equation}
and
\begin{equation}
d\Omega _{1\bar{2}}=-3\frac{\chi V_{I}X^{I}}{f^{4}}.
\end{equation}

The constraints ({\ref{nextsim12}}) and ({\ref{nextsim13}}) then imply that

\begin{equation}
d\Omega _{1\bar{1}}=d\Omega _{2\bar{2}}=0
\end{equation}
together with
\begin{equation}
d\Omega _{12}=-{\frac{i}{\sqrt{2}}}(\frac{1}{T}{\frac{\partial \mathcal{G}} {%
\partial x^{2}}}+\frac{\mathcal{H}}{T^{2}}{\frac{\partial T}{\partial x^{1}}}%
)
\end{equation}
and the constraints
\begin{equation}
{\frac{\partial }{\partial x^{2}}}(T\mathcal{H})={\frac{\partial }{\partial
x^{1}}}(T\mathcal{G}),  \label{difconstr1v2}
\end{equation}
and

\begin{equation}
{\frac{\partial }{\partial x^{1}}}\left( \frac{\mathcal{H}}{T}\right) ={%
\frac{\partial }{\partial x^{2}}}\left( \frac{\mathcal{G}}{T}\right) ,\qquad
{\frac{\partial }{\partial x^{1}}}\left( \frac{\mathcal{G}}{T}\right) =-{%
\frac{\partial }{\partial x^{2}}}\left( \frac{\mathcal{H}}{T}\right) .
\label{difconstr2v2}
\end{equation}%
Just as in the previous section, these constraints imply that $T^{-1}%
\mathcal{H}$ and $T^{-1}\mathcal{G}$ satisfy the Cauchy-Riemann equations; $%
T^{-1}(\mathcal{H}+i\mathcal{G})$ is a holomorphic function of $x^{1}+ix^{2}$%
.

In these co-ordinates $d\Omega $ takes the form

\begin{eqnarray}
d\Omega &=&-{\frac{i}{\sqrt{2}}}(T{\frac{\partial \mathcal{G}}{\partial x^{2}%
}}+\mathcal{H}{\frac{\partial T}{\partial x^{1}}})\left( dx^{1}\wedge dx^{2}-%
\frac{1}{T^{2}}d\phi \wedge d\psi \right)  \notag \\
&-&\frac{3\chi }{f^{4}}V_{I}X^{I}\left( T^{2}dx^{1}\wedge dx^{2}+d\phi
\wedge d\psi \right) .  \label{domegsol3}
\end{eqnarray}

The gauge field strengths for these solutions can be written as
\begin{equation}
F^{I}=d\left( f^{2}X^{I}(dt+\Omega )\right) +6\chi \frac{T^{2}}{f^{2}}%
V_{J}(X^{I}X^{J}-{\frac{1}{2}}Q^{IJ})dx^{1}\wedge dx^{2}
\end{equation}%
which automatically satisfy the Bianchi identity $dF^{I}=0$.

\bigskip Finally the integrability condition associated with ({\ref%
{eqn:killpot1}}) can be written as

\begin{equation}
\Box \log T=18\frac{\chi ^{2}}{f^{2}}(X^{I}X^{J}-{\frac{1}{2}}
Q^{IJ})V_{I}V_{J}T^{2}.
\end{equation}
It is then straightforward to show that these constraints imply that the
integrability condition $d^{2}\Omega =0$ associated with the expression in ({%
\ref{domegsol3}}) holds automatically. Lastly, the constraints ({\ref%
{nextsimp10}}) and ({\ref{nextsimp11}}) fix $d(\func{Re}\frac{\lambda }{f})$
and $d(\func{Re}\frac{\sigma }{f})$ in terms of constant linear combinations
of $d\phi $ and $d\psi $; these conditions do not impose any further
constraints on the geometry.

\section{Summary and Discussion}

In this paper we have employed the spinorial geometry method for the task of
classifying $1/2$ supersymmetric solutions with at least one time-like
Killing spinor of the theory of $N=2,D=5$ supergravity. Our results provide
a general framework for the explicit construction of many new black holes
and the investigation of their physical properties and relevance to AdS/CFT\
correspondence and holography.

In general, supersymmetric solutions in five dimensional theories must
preserve either $2$, $4$, $6$ or $8$ of the supersymmetries. This is because
the Killing spinor equations are linear over $\mathbb{C}$. However in the
ungauged theories, it was found that supersymmetric solutions can only
preserve $4$ or $8$ of supersymmetries \cite{pakis, unique}. Moreover, to
find time-like supersymmetric solutions in the ungauged theory, one must
solve the gauge equations and the Bianchi identities in addition to the
Killing spinor equations. In the null case one must additionally solve one
of the components of the Einstein equations of motion. A similar situation
arises for $1/4$ supersymmetric solutions in the gauged theory. However, for
the solutions with $1/2$ supersymmetry considered in our present work, we
have demonstrated that if one of the Killing spinors is time-like, then
supersymmetry and Bianchi identities alone imply that all components of
Einstein and gauge equations together with the scalar equations are
automatically satisfied.

Maximally supersymmetric solutions (preserving all 8 of the supersymmetries)
of the five dimensional gauged supergravity theory have vanishing gauge
field strengths and constant scalars and are locally isometric to $AdS_{5}$.
Moreover, it has been demonstrated in \cite{preons} that all solutions of $%
N=2$, $D=5$ supergravity preserving $3/4$ of supersymmetry must be locally
isometric to $AdS_5$, with vanishing gauge field strengths and constant
scalars. An analogous situation also arises in the case of $D=11$
supergravity, where it has been shown that all solutions with $31/32$
supersymmetry must be locally isometric to a maximally supersymmetric
solution \cite{d11preons}. However, in the case of $D=11$ supergravity, it
has been shown that one cannot obtain $31/32$ supersymmetric solutions by
taking quotients of the maximally supersymmetric solutions \cite%
{josenopreons}. In contrast, there is a $3/4$-supersymmetric supersymmetric
solution of $N=2$, $D=5$ gauged supergravity which is obtained by taking a
certain quotient of $AdS_5$ \cite{josenopreons2}.

One future direction is the completion of the classification of
supersymmetric solutions in $N=2$, $D=5$ supergravity by classifying $1/2$
supersymmetric solutions with two null Killing spinors (i.e., two Killing
spinors with associated null Killing vectors). In addition, it would be
interesting to investigate whether there are any regular asymptotically $%
AdS_5$ black ring solutions (see \cite{har} for a recent discussion).
Supersymmetric black rings are known to exist in the ungauged theory \cite%
{harveyelv:04, harveyelv:04b, ggbr1:04, ggbr2:04}. For asymptotically flat
supersymmetric black rings of $N=2$, $D=5$ ungauged supergravity,
supersymmetry is enhanced from $1/2$ supersymmetry to maximal supersymmetry
at the horizon. If there do exist $AdS_5$ black rings in the gauged theory,
it may be reasonable to expect that the supersymmetry will be enhanced from $%
1/4$ to $1/2$ at the horizon. We hope that the classification of $1/2$
supersymmetric solutions can provide a method of determining whether there
exists a $1/2$ supersymmetric solution corresponding to the near-horizon
geometry of a black ring. It should be noted that in \cite{har}, black rings
with two $U(1)$ symmetries have already been excluded. However, as we have
seen, $1/2$ supersymmetric solutions in general only have one additional $%
U(1)$ symmetry; so more general ring solutions may be possible.

\acknowledgments

W. Sabra would like to thank Queen Mary University of London and Khuri Lab
at Rockefeller University for hospitality during which some of this work was
completed. The work of W. Sabra was supported in part by the National
Science Foundation under grant number PHY-0601213.

\medskip \vfill

\appendix

\section{Systematic treatment of the Killing spinor equation}

In this appendix we will evaluate the linear system obtained from the
Killing spinor acting on $\eta ^{a}$ given in ({\ref{sali}}), keeping the
parameters arbitrary. It would naively appear that we have to evaluate two
sets of Killing spinor equations, according to the two choices of symplectic
index $a$. However, making use of the symplectic Majorana condition,
together with the fact that the gauge field strengths and scalars are real,
it is straightforward to show that it suffices just to consider the case
when $a=1$, the $a=2$ equations are then implied automatically. In the
following, it will be convenient to define $H=X_{I}F^{I}$ and $\mu _{\bar{p}%
}=\delta _{\bar{p}q}\mu ^{q}$.

From the dilatino equation we obtain

\begin{eqnarray}
&&\sigma \left( F^{I}{}_{mn}-X^{I}H_{mn}\right) \epsilon^{mn}-\lambda \left(
F^{I}{}_{m}{}^{m}-X^{I}H_{m}{}^{m}+\partial_{0}X^{I}\right) =  \notag \\
&&2\chi V_{J}(X^{I}X^{J}-{\frac{3}{2}}Q^{IJ})\sigma ^{\ast }+\sqrt{2}i\left(
F^{I}{}_{0m}-X^{I}H_{0m}-\partial_{m}X^{I}\right) \mu^{m}  \notag \\
&&  \label{alg} \\
&&\sqrt{2}i\sigma \left( F^{I}{}_{0m} \epsilon^{m}{}_{\bar{q}%
}-X^{I}H{}_{0m}+\partial_{m}X^{I}\right) \epsilon ^{m}{}_{\bar{q}}- \mu _{%
\bar{q}}\left( F^{I}{}_{m}{}^{m}-X^{I}H{}_{m}{}^{m}-
\partial_{0}X^{I}\right) =  \notag \\
&&2\chi V_{J}(X^{I}X^{J}-{\frac{3}{2}}Q^{IJ})\epsilon _{\bar{m}\bar{q}}
(\mu^{m})^{\ast }-\sqrt{2}i\lambda \left( F^{I}{}_{0\bar{q}}-X^{I}H{}_{0\bar{%
q} }+\partial _{\bar{q}}X^{I}\right)  \notag \\
&&-2\left( F^{I}{}_{m\bar{q}}-X^{I}H{}_{m\bar{q}}\right) \mu^{m}  \notag \\
&& \\
&&\sqrt{2}i\left( -F^{I}{}_{0\bar{m}}+X^{I}H{}_{0\bar{m}}+ \partial_{\bar{m}%
}X^{I}\right) \epsilon^{\bar{m}}{}_{n}\mu^{n} -\lambda \epsilon^{\bar{m}
\bar{n}}\left( F^{I}{}_{\bar{m}\bar{n}}-X^{I}H{}_{\bar{m}\bar{n}}\right)
\notag \\
&=&-2\chi V_{J}(X^{I}X^{J}-{\frac{3}{2}}Q^{IJ})\lambda^{\ast }-\sigma \left(
F^{I}{}_{m}{}^{m}-\partial _{0}X^{I}-X^{I}H{}_{m}{}^{m}\right)  \notag \\
&&
\end{eqnarray}
Then from the gravitino part of the Killing spinor equations we obtain the
following constraints:

From along the $0$-direction of the supercovariant derivative-

\begin{eqnarray}
\partial _{0}\lambda &=&-\frac{i\mu ^{m}}{\sqrt{2}}\left( -\omega
_{0,0m}+H_{0m}\right) -{\frac{1}{4}}\sigma \left( 2\omega
_{0,mn}+H_{mn}\right) \epsilon ^{mn}  \notag \\
&&+{\frac{1}{4}}\lambda \left( H_{m}{}^{m}+2\omega_{0,m}{}^{m}\right) -{%
\frac{\chi }{2}}V_{I}\left( X^{I}-{3}A_{0}^{I}\right) \sigma^{\ast },
\end{eqnarray}

\begin{eqnarray}
\partial _{0}\mu _{\bar{q}} &=&-\frac{i\sigma }{\sqrt{2}}\left( \omega
_{0,0m}\epsilon ^{m}{}_{\bar{q}}+H_{0m}\epsilon ^{m}{}_{\bar{q}}\right) -{%
\frac{i}{\sqrt{2}}}\lambda \left( \omega _{0,0\bar{q}}+H_{0\bar{q}}\right) -%
\frac{1}{4}\left( H_{m}{}^{m}-2\omega _{0,m}{}^{m}\right) \mu _{\bar{q}}
\notag \\
&&-\left( -{\frac{1}{2}}H_{m\bar{q}}+\omega _{0,m\bar{q}}\right) \mu ^{m} +%
\frac{\chi }{2}V_{I}\left( X^{I}+{3}A_{0}^{I}\right) \epsilon _{\bar{m} \bar{%
q}}(\mu ^{m})^{\ast },
\end{eqnarray}

\begin{eqnarray}
\partial _{0}\sigma &=&-\frac{i}{\sqrt{2}}\left( -\omega _{0,0\bar{m}} +H_{0%
\bar{m}}\right) \epsilon ^{\bar{m}}{}_{n}\mu ^{n}-{\frac{1}{4}}\sigma
\left(2\omega _{0,m}{}^{m}+H_{m}{}^{m}\right)  \notag \\
&&+{\frac{1}{4}}\lambda \left( H_{\bar{m}\bar{n}} +2\omega _{0,\bar{m}\bar{n}%
}\right) \epsilon ^{\bar{m}\bar{n}} +\frac{\chi }{2}V_{I}(X^{I}-{3}%
A_{0}^{I})\lambda ^{\ast }.
\end{eqnarray}
From along the $p$-direction of the supercovariant derivative-

\begin{equation}
\partial _{p}\lambda =\frac{i}{\sqrt{2}}\left( \omega _{p,0m}-{\frac{3}{2}}
H_{pm}\right) \mu ^{m}-\frac{\sigma }{2}\omega _{p,mn}\epsilon ^{mn} +\frac{%
\lambda }{2}\left( \omega _{p,m}{}^{m}-{\frac{3}{2}}H_{0p}\right) +{\frac{%
3\chi }{2}}V_{I}A_{p}^{I}\sigma ^{\ast },
\end{equation}

\begin{eqnarray}
\partial _{p}\mu _{\bar{q}} &=&\frac{i\sigma }{2\sqrt{2}}\left( -2\omega
_{p,0m}\epsilon ^{m}{}_{\bar{q}}+{\frac{1}{2}}H_{mn}\epsilon ^{mn} \delta _{p%
\bar{q}}\right) -\frac{i\lambda }{2\sqrt{2}}\left( 2\omega _{p,0\bar{q}} +{3}%
H_{p\bar{q}}-H_{m}{}^{m}\delta _{p\bar{q}}\right)  \notag \\
&&+\left( \frac{1}{2}\omega _{p,m}{}^{m}+{\frac{3}{4}}H_{0p}\right) \mu _{%
\bar{q}}-\mu ^{m}\left( \omega _{p,m\bar{q}}+{\frac{1}{2}}H_{0m} \delta _{p%
\bar{q}}\right)  \notag \\
&&+\chi V_{I}\left( -{\frac{i}{\sqrt{2}}}X^{I}\sigma ^{\ast } \delta _{p\bar{%
q}}+{\frac{3}{2}}A_{p}^{I}\epsilon _{\bar{m}\bar{q}} (\mu ^{m})^{\ast
}\right),
\end{eqnarray}

\begin{eqnarray}
\partial _{p}\sigma &=&\frac{i}{2\sqrt{2}}\left( 2\omega _{p,0\bar{m}} -H_{p%
\bar{m}}\right) \epsilon ^{\bar{m}}{}_{n}\mu ^{n}+\frac{\lambda }{2}
\left(\omega _{p,\bar{m}\bar{n}}\epsilon ^{\bar{m}\bar{n}}-H_{0\bar{n}}
\epsilon ^{\bar{n}}{}_{p}\right) -\frac{\sigma }{4}\left(
2\omega_{p,m}{}^{m}+H_{0p}\right)  \notag \\
&&-{\frac{i}{2\sqrt{2}}}H_{m}{}^{m}\epsilon _{p}{}^{\bar{n}} \mu _{\bar{n}%
}-\chi V_{I}\left( {\frac{i}{\sqrt{2}}}X^{I}(\mu _{\bar{p}})^{\ast } +{\frac{%
3}{2}}A_{p}^{I}\lambda ^{\ast }\right) .
\end{eqnarray}
From along the ${\bar{p}}$-direction of the supercovariant derivative-

\begin{eqnarray}
\partial _{\bar{p}}\lambda &=&\frac{i}{2\sqrt{2}}\left( 2\omega _{{\bar{p}}%
,0m}-H_{\bar{p}m}\right) \mu ^{m}+{\frac{\lambda }{4}}\left( 2\omega _{{\bar{%
p}},m}{}^{m}-H_{0\bar{p}}\right) +\frac{1}{2}\sigma \left( H_{0m}\epsilon
^{m}{}_{\bar{p}}-\omega _{{\bar{p}},mn}\epsilon ^{mn}\right)  \notag \\
&&+{\frac{i}{2\sqrt{2}}}H_{m}{}^{m}\mu _{\bar{p}}+\chi V_{I}\left( {\frac{i}{%
\sqrt{2}}}X^{I}\epsilon _{\bar{p}\bar{m}}(\mu ^{m})^{\ast }+{\frac{3}{2}}A_{%
\bar{p}}^{I}\sigma ^{\ast }\right) ,
\end{eqnarray}

\begin{eqnarray}
\partial _{\bar{p}}\mu _{\bar{q}} &=& -{\frac{i}{\sqrt{2}}} \lambda \left(
\omega_{{\bar{p}},0 {\bar{q}}}+{\frac{1}{4}} H_{\bar{m} \bar{n}} \epsilon^{%
\bar{m} \bar{n}} \epsilon_{\bar{p} \bar{q}} \right) -{\frac{i}{\sqrt{2}}}
\sigma \left(\omega_{\bar{p},0m} \epsilon^m{}_{\bar{q}} + {\frac{1}{2}}
H_m{}^m \epsilon_{\bar{p} \bar{q}} +{\frac{3}{2}} H_{\bar{p} m}\epsilon^m{}_{%
\bar{q}} \right)  \notag \\
&-& \mu^n \left( \omega_{\bar{p},n \bar{q}} +{\frac{1}{2}} H_{0 \bar{m}}
\epsilon^{\bar{m}}{}_n \epsilon_{\bar{p} \bar{q}} \right) +\left( {\frac{3}{4%
}}H_{0\bar{p}}+\frac{1}{2}\omega _{{\bar{p}},m}{}^{m}\right) \mu _{\bar{q}}
\notag \\
&+&\chi V_{I}\left( {\frac{i}{\sqrt{2}}} X^{I}\lambda ^{\ast }\epsilon _{%
\bar{p}\bar{q}}+{\frac{3}{2}} A_{\bar{p}}^{I}\epsilon _{\bar{m}\bar{q}}(\mu
^{m})^{\ast }\right) ,
\end{eqnarray}

\begin{eqnarray}
\partial _{\bar{p}}\sigma &=&\frac{i}{\sqrt{2}} \omega _{{\bar{p}},0\bar{m}%
}\epsilon ^{\bar{m}}{}_{n}\mu ^{n}+\frac{1}{2}\lambda \omega _{{\bar{p}},%
\bar{m}\bar{n}}\epsilon ^{\bar{m}\bar{n}}-\sigma \left( {\frac{1}{2}} \omega
_{{\bar{p}},m}{}^{m}+{\frac{3}{4}}H_{0\bar{p}}\right)  \notag \\
&&+{\frac{3i}{4\sqrt{2}}}H_{\bar{m}\bar{n}}\epsilon ^{\bar{m}\bar{n}} \mu _{%
\bar{p}}-{\frac{3\chi }{2}}V_{I}A_{\bar{p}}^{I}\lambda ^{\ast }.
\label{algg}
\end{eqnarray}

Throughout these equations, spatial indices of $\epsilon _{mn}$, $\epsilon _{%
\bar{m}\bar{n}}$, $\omega _{A,BC}$, $F^{I}{}_{AB}$ and $H_{AB}$ have been
raised with $\delta ^{p\bar{q}}$.

\section{Integrability Conditions and Equations of Motion}

In this appendix we examine the integrability conditions of the Killing
spinor equations. It will be shown that, if a background preserves at least
half of the supersymmetry, and admits a Killing spinor for which the
associated Killing vector is time-like, and the Bianchi identity holds, then
all components of the Einstein, gauge and scalar equations hold
automatically.

First we consider the integrability condition associated with the gravitino
equation ({\ref{eqn:grav}}). After some gamma matrix manipulation we find

\begin{eqnarray}  \label{eqn:gravintv1}
0 &=&-{\frac{1}{4}}R_{\alpha \beta \beta _{1}\beta _{2}}\gamma ^{\beta
_{1}\beta _{2}}\epsilon ^{a}  \notag \\
&&-{\frac{1}{4}}\nabla _{\lbrack \alpha }X_{I}(\gamma _{\beta ]}{}^{\beta
_{1}\beta _{2}}-4\delta _{\beta ]}^{\beta _{1}}\gamma ^{\beta
_{2}})F^{I}{}_{\beta _{1}\beta _{2}}\epsilon ^{a}  \notag \\
&&+{\frac{1}{4}}X_{I}(\gamma _{\lbrack \alpha }{}^{\beta _{1}\beta
_{2}}-4\delta _{\lbrack \alpha }^{\beta _{1}}\gamma ^{\beta _{2}})\nabla
_{\beta ]}F^{I}{}_{\beta _{1}\beta _{2}}\epsilon ^{a}  \notag \\
&&+\chi V_{I}\left( \nabla _{\lbrack \alpha }X^{I}\gamma _{\beta ]}- {\frac{3%
}{2}}F^{I}{}_{\alpha \beta }\right) \epsilon ^{ab}\epsilon ^{b}  \notag \\
&&+{\frac{1}{4}}X_{I}X_{J}\big(F^{I}{}_{\beta _{1}\beta _{2}}F^{J}{}_{\beta
_{3}[\alpha }\gamma _{\beta ]}{}^{\beta _{1}\beta _{2}\beta
_{3}}+F^{I}{}_{\beta _{1}[\alpha }F^{J\beta _{1}\beta _{2}}\gamma _{\beta
]\beta _{2}}  \notag \\
&&+{\frac{1}{4}}F^{I}{}_{\beta _{1}\beta _{2}}F^{J\beta _{1}\beta
_{2}}\gamma _{\alpha \beta }-{\frac{3}{2}}F^{I}{}_{\alpha \beta
_{1}}F^{J}{}_{\beta \beta _{2}}\gamma ^{\beta _{1}\beta _{2}}\big)\epsilon
^{a}  \notag \\
&&+{\frac{\chi }{4}}V_{I}X^{I}X_{J}\big(F^{J}{}_{\beta _{1}\beta _{2}}\gamma
_{\alpha \beta }{}^{\beta _{1}\beta _{2}}+4\gamma ^{\mu }{}_{[\alpha
}F^{J}{}_{\beta ]\mu }\big)\epsilon ^{ab}\epsilon ^{b}  \notag \\
&&+{\frac{\chi ^{2}}{2}}V_{I}V_{J}X^{I}X^{J}\gamma _{\alpha \beta }\epsilon
^{a}.
\end{eqnarray}

Next consider the dilatino equation ({\ref{eqn:newdil}}). This gives the
integrability condition

\begin{eqnarray}  \label{eqn:dilatintv1}
0 &=&{\frac{3}{4}}\gamma ^{\beta }\nabla _{\alpha } \nabla
_{\beta}X_{I}\epsilon ^{a}  \notag \\
&&+{\frac{3\chi }{2}}\big(\nabla _{\alpha }(X_{I}V_{J}X^{J})+{\frac{1}{2}}
V_{J}X^{J}\nabla _{\beta }X_{I}\gamma ^{\beta }{}_{\alpha }\big)\epsilon
^{ab}\epsilon ^{b}  \notag \\
&&+\nabla _{\alpha }\big(({\frac{1}{4}}Q_{IJ}-{\frac{3}{8}}
X_{I}X_{J})F^{J}{}_{\beta _{1}\beta _{2}}\big) \gamma ^{\beta
_{1}\beta_{2}}\epsilon ^{a}  \notag \\
&&+\big({\frac{3}{16}}X_{J}F^{J}{}_{\beta _{1}\beta _{2}} \nabla
_{\beta_{3}}X_{I}\gamma _{\alpha }{}^{\beta _{1}\beta _{2}\beta _{3}}-{\frac{%
3}{4}} X_{J}F^{J}{}_{\alpha \beta _{1}}\nabla _{\beta _{2}}X_{I}\gamma
^{\beta _{1}\beta _{2}}\big)\epsilon ^{a}  \notag \\
&&-2\chi V_{K}X^{K}({\frac{1}{4}}Q_{IJ}-{\frac{3}{8}}X_{I}X_{J})
F^{J}{}_{\alpha \beta }\gamma ^{\beta }\epsilon ^{ab}\epsilon ^{b}  \notag \\
&&+X_{K}\big(({\frac{1}{8}}Q_{IJ}-{\frac{3}{16}}X_{I}X_{J})F^{J}{}_{\alpha
\beta _{1}}F^{K}{}_{\beta _{2}\beta _{3}}\gamma ^{\beta _{1}\beta _{2}\beta
_{3}}  \notag \\
&&-{\frac{1}{16}}C_{IJM}X^{M}F^{J}{}_{\beta _{1}\mu }F^{K}{}_{\beta
_{2}}{}^{\mu }\gamma _{\alpha }{}^{\beta _{1}\beta _{2}}+({\frac{1}{2}}
Q_{IJ}-{\frac{3}{4}}X_{I}X_{J})F^{J}{}_{\beta \mu } F^{K}{}_{\alpha
}{}^{\mu}\gamma ^{\beta }\big)\epsilon ^{a} \ .  \notag \\
&&  \label{dilatint}
\end{eqnarray}

It will be convenient to define

\begin{eqnarray}
E_{\alpha \beta } &=&R_{\alpha \beta }+Q_{IJ}F^{I}{}_{\alpha \mu
}F^{J}{}_{\beta }{}^{\mu }-Q_{IJ}\nabla _{\alpha }X^{I}\nabla _{\beta }X^{J}
\notag \\
&&+g_{\alpha \beta }\left( -{\frac{1}{6}}Q_{IJ} F^{I}{}_{\beta
_{1}\beta_{2}}F^{J\beta _{1}\beta _{2}}+6\chi ^{2}({\frac{1}{2}}
Q^{IJ}-X^{I}X^{J})V_{I}V_{J}\right)  \notag \\
G_{I\alpha } &=&\nabla ^{\beta }\left( Q_{IJ}F^{J}{}_{\alpha \beta }\right)
+ {\frac{1}{16}}C_{IJK}\epsilon _{\alpha }{}^{\beta _{1}\beta _{2}\beta
_{3}\beta _{4}}F^{J}{}_{\beta _{1}\beta _{2}}F^{K}{}_{\beta _{3}\beta _{4}}
\notag \\
S_{I} &=&\nabla ^{\alpha }\nabla _{\alpha }X_{I}-({\frac{1}{6}}C_{MNI}- {%
\frac{1}{2}}X_{I}C_{MNJ}X^{J})\nabla _{\alpha }X^{M}\nabla ^{\alpha }X^{N}
\notag \\
&&-{\frac{1}{2}}\left( X_{M}X^{P}C_{NPI}-{\frac{1}{6}}
C_{MNI}-6X_{I}X_{M}X_{N}+{\frac{1}{6}}X_{I}C_{MNJ}X^{J}\right)
F^{M}{}_{\beta _{1}\beta _{2}}F^{N\beta _{1}\beta _{2}}  \notag \\
&&-3\chi ^{2}V_{M}V_{N}\left( {\frac{1}{2}}
Q^{ML}Q^{NP}C_{LPI}+X_{I}(Q^{MN}-2X^{M}X^{N})\right)
\end{eqnarray}
so that $E_{\alpha \beta }=0$, $G_{I\alpha }=0$ and $S_{I}=0$ correspond to
the Einstein, gauge field and scalar equations of motion respectively.

To proceed we act on the gravitino integrability condition ({\ref%
{eqn:gravintv1}}) from the left with $\gamma^{\beta}$ and contract over the
index $\beta$. We assume that the Bianchi identity $dF^I=0$ holds. After
some considerable gamma matrix manipulation (and making use of ({\ref%
{eqn:newdil}}) to simplify the expressions further), we find the constraint

\begin{equation}
\left( E_{\alpha \beta }\gamma ^{\beta }+\frac{1}{3}X^{I}\left( \gamma
_{\alpha }{}^{\beta }G_{I\beta }-{2}G_{I\alpha }\right) \right) \epsilon
^{a}=0.  \label{eqn:gravinteq}
\end{equation}
Also, on contracting the dilatino integrability condition ({\ref%
{eqn:dilatintv1}}) with $\gamma ^{\alpha }$, and again assuming the Bianchi
identity $dF^{I}=0$ holds, we find

\begin{equation}
\left( S_{I}-{\frac{2}{3}}(G_{I\alpha }-X_{I}X^{J}G_{J\alpha })\gamma
^{\alpha }\right) \epsilon ^{a}=0.  \label{eqn:scalinteq}
\end{equation}

To proceed, we evaluate the constraints ({\ref{eqn:gravinteq}}) and ({\ref%
{eqn:scalinteq}}) on a background which preserves at least half of the
supersymmetry, and which admits a Killing spinor for which the associated
Killing vector is time-like. In particular, we first consider a generic
Killing spinor

\begin{eqnarray}
\eta ^{1} &=&\lambda 1+\mu ^{i}e^{i}+\sigma e^{12}, \\
\eta ^{2} &=&-\sigma ^{\ast }1-\epsilon _{ij}(\mu ^{i})^{\ast }e^{j}+\lambda
^{\ast }e^{12}.
\end{eqnarray}%
Substituting this expression into ({\ref{eqn:gravinteq}}) for $\alpha =0$
gives the constraints

\begin{eqnarray}
\lambda E_{00}-\sqrt{2}i\mu ^{p}\left( E_{0p}+{\frac{1}{3}}%
X^{I}G_{Ip}\right) -{\frac{2}{3}}\lambda X^{I}G_{I0} &=&0,  \notag \\
E_{00}\mu ^{p}+\sqrt{2}i\sigma \left( E_{0q}-{\frac{1}{3}}X^{I}G_{Iq}\right)
\epsilon ^{qp}+\sqrt{2}i\lambda \left( E_{0}{}^{p}-{\frac{1}{3}}%
X^{I}G_{I}{}^{p}\right) +{\frac{2}{3}}X^{I}G_{I0}\mu ^{p} &=&0,  \notag \\
\sigma E_{00}-\sqrt{2}i\left( E_{0\bar{p}}+{\frac{1}{3}}X^{I}G_{I\bar{p}%
}\right) \epsilon ^{\bar{p}}{}_{q}\mu ^{q}-{\frac{2}{3}}X^{I}G_{I0}\sigma
&=&0.  \notag \\
&&  \label{gravin1}
\end{eqnarray}%
Evaluating these constraints on the canonical form of the $N=1$ time-like
Killing spinor by setting $\lambda =f,\mu ^{1}=\mu ^{2}=\sigma =0$ we obtain
the constraints

\begin{eqnarray}
E_{00} &=&{\frac{2}{3}}X^{I}G_{I0},  \notag \\
E_{0p} &=&{\frac{1}{3}}X^{I}G_{Ip},  \notag \\
E_{0\bar{p}} &=&{\frac{1}{3}}X^{I}G_{I\bar{p}}.
\end{eqnarray}
Now substitute these expressions back into (\ref{gravin1}) and eliminate the
$E_{\alpha \beta }$ terms to find

\begin{eqnarray}
X^{I}G_{Ip}\mu ^{p} &=&0,  \notag \\
X^{I}G_{I0}\mu ^{p} &=&0,  \notag \\
X^{I}G_{I\bar{p}}\epsilon ^{\bar{p}}{}_{q}\mu ^{q} &=&0.
\end{eqnarray}
Assuming that the background is at least half-supersymmetric, we take $(\mu
^{1},\mu ^{2})\neq (0,0)$, and hence find

\begin{equation}
X^{I}G_{I\alpha }=0
\end{equation}

and

\begin{equation}
E_{00}=E_{0p}=0.
\end{equation}
Next consider ({\ref{eqn:gravinteq}}) for $\alpha =p$ evaluated on the
generic spinor $\eta ^{1}$. Using the constraints $E_{0p}=0$ and $%
X^{I}G_{I\alpha }=0$ which we have already obtained, this expression
simplifies to

\begin{equation}
(E_{pq}\gamma ^{q}+E_{p\bar{q}}\gamma ^{\bar{q}})\eta ^{1}=0
\end{equation}
from which we find the constraints

\begin{eqnarray}
E_{pq}\mu ^{q} &=&0,  \notag \\
\sigma E_{pq}\epsilon ^{q}{}_{\bar{\ell}}+\lambda E_{p\bar{\ell}} &=&0,
\notag \\
E_{p\bar{q}}\epsilon ^{\bar{q}}{}_{\ell }\mu ^{\ell } &=&0
\end{eqnarray}
and taking ({\ref{eqn:gravinteq}}) with $\alpha =p$ we find

\begin{eqnarray}
E_{\bar{p}q}\mu ^{q} &=&0,  \notag \\
\sigma E_{\bar{p}q}\epsilon ^{q}{}_{\bar{\ell}}+\lambda E_{\bar{p}\bar{\ell}%
} &=&0,  \notag \\
E_{\bar{p}\bar{q}}\epsilon ^{\bar{q}}{}_{\ell }\mu ^{\ell } &=&0.
\end{eqnarray}

Evaluating these constraints on the canonical $N=1$ time-like spinor by
taking $\lambda =f,\mu ^{1}=\mu ^{2}=\sigma =0,$ we obtain the constraints

\begin{equation}
E_{p\bar{q}}=0,\quad E_{pq}=0.
\end{equation}
Hence we have shown that for solutions with at least half supersymmetry, the
constraint ({\ref{eqn:gravinteq}}) implies that

\begin{equation}
E_{\alpha \beta }=0,\quad X^{I}G_{I\alpha }=0.
\end{equation}

Next consider the constraint ({\ref{eqn:scalinteq}}) obtained from the
dilatino integrability conditions. On using $X^I G_{I \alpha}=0$ this
constraint simplifies to

\begin{equation}
\left( S_{I}-{\frac{2}{3}}G_{I\alpha }\gamma ^{\alpha }\right) \epsilon
^{a}=0.
\end{equation}

Evaluating this expression on the generic Killing spinor $\eta ^{a}$, one
obtains

\begin{eqnarray}
\lambda \left( S_{I}-{\frac{2}{3}}G_{I0}\right) +{\frac{2\sqrt{2}i}{3}}%
G_{Ip}\mu ^{p} &=&0,  \notag \\
S_{I}\mu ^{p}+{\frac{2}{3}}G_{I0}\mu ^{p}+{\frac{2\sqrt{2}i}{3}}\left(
\sigma G_{Iq}\epsilon ^{qp}+\lambda G_{I}{}^{p}\right) &=&0,  \notag \\
\sigma \left( S_{I}-{\frac{2}{3}}G_{I0}\right) +{\frac{2\sqrt{2}i}{3}}G_{I%
\bar{p}}\epsilon ^{\bar{p}}{}_{q}\mu ^{q} &=&0.  \label{safso}
\end{eqnarray}%
Evaluating these constraints on the canonical $N=1$ time-like spinor by
taking $\lambda =f,\mu ^{1}=\mu ^{2}=\sigma =0$, the following conditions
are obtained

\begin{eqnarray}
S_{I} &=&{\frac{2}{3}}G_{I0},  \notag \\
G_{Ip}=G_{I\bar{p}} &=&0.
\end{eqnarray}
Now substitute these constraints back into ({\ref{safso}}) to find

\begin{equation}
G_{I0}\mu ^{p}=0.
\end{equation}
Assuming that the background is at least half-supersymmetric, we take $(\mu
^{1},\mu ^{2})\neq (0,0)$, and hence

\begin{equation}
G_{I0}=0.
\end{equation}
Hence from the constraint ({\ref{eqn:scalinteq}}) we have found the
constraints

\begin{equation}
S_{I}=0,\quad G_{I\alpha }=0.
\end{equation}

\bigskip \vfill\eject

\end{document}